\documentclass[longauth]{aa}  

\usepackage{graphicx}
\usepackage{subfigure}
\usepackage{adjustbox}
\usepackage{txfonts}
\usepackage{natbib}
\usepackage{multirow}
\bibpunct{(}{)}{;}{a}{}{,} 
\usepackage[colorlinks=true,linkcolor=blue,allcolors=blue]{hyperref}
\begin{document} 

\title{Long-term multi-wavelength study of 1ES\,0647+250}


\author{\small
MAGIC Collaboration: 
V.~A.~Acciari\inst{1} \and
T.~Aniello\inst{2} \and
S.~Ansoldi\inst{3,45} \and
L.~A.~Antonelli\inst{2} \and
A.~Arbet Engels\inst{4} \and
C.~Arcaro\inst{5} \and
M.~Artero\inst{6} \and
K.~Asano\inst{7} \and
D.~Baack\inst{8} \and
A.~Babi\'c\inst{9}
% ...
}

\institute{Instituto de Astrof\'isica de Canarias and Dpto. de  Astrof\'isica, Universidad de La Laguna, E-38200, La Laguna, Tenerife, Spain
\and National Institute for Astrophysics (INAF), I-00136 Rome, Italy
\and Universit\`a di Udine and INFN Trieste, I-33100 Udine, Italy
\and Max-Planck-Institut f\"ur Physik, D-80805 M\"unchen, Germany
\and Universit\`a di Padova and INFN, I-35131 Padova, Italy
\and Institut de F\'isica d'Altes Energies (IFAE), The Barcelona Institute of Science and Technology (BIST), E-08193 Bellaterra (Barcelona), Spain
\and Japanese MAGIC Group: Institute for Cosmic Ray Research (ICRR), The University of Tokyo, Kashiwa, 277-8582 Chiba, Japan
\and Technische Universit\"at Dortmund, D-44221 Dortmund, Germany
\and Croatian MAGIC Group: University of Zagreb, Faculty of Electrical Engineering and Computing (FER), 10000 Zagreb, Croatia
\and IPARCOS Institute and EMFTEL Department, Universidad Complutense de Madrid, E-28040 Madrid, Spain
\and Centro Brasileiro de Pesquisas F\'isicas (CBPF), 22290-180 URCA, Rio de Janeiro (RJ), Brazil
\and University of Lodz, Faculty of Physics and Applied Informatics, Department of Astrophysics, 90-236 Lodz, Poland
\and Instituto de Astrof\'isica de Andaluc\'ia-CSIC, Glorieta de la Astronom\'ia s/n, 18008, Granada, Spain
\and Deutsches Elektronen-Synchrotron (DESY), D-15738 Zeuthen, Germany
\and ETH Z\"urich, CH-8093 Z\"urich, Switzerland
\and Universit\`a di Pisa and INFN Pisa, I-56126 Pisa, Italy
\and Universitat de Barcelona, ICCUB, IEEC-UB, E-08028 Barcelona, Spain
\and Armenian MAGIC Group: A. Alikhanyan National Science Laboratory, 0036 Yerevan, Armenia
\and Centro de Investigaciones Energ\'eticas, Medioambientales y Tecnol\'ogicas, E-28040 Madrid, Spain
\and Department for Physics and Technology, University of Bergen, Norway
\and INFN MAGIC Group: INFN Sezione di Catania and Dipartimento di Fisica e Astronomia, University of Catania, I-95123 Catania, Italy
\and INFN MAGIC Group: INFN Sezione di Torino and Universit\`a degli Studi di Torino, I-10125 Torino, Italy
\and INFN MAGIC Group: INFN Sezione di Bari and Dipartimento Interateneo di Fisica dell'Universit\`a e del Politecnico di Bari, I-70125 Bari, Italy
\and Croatian MAGIC Group: University of Rijeka, Department of Physics, 51000 Rijeka, Croatia
\and Universit\"at W\"urzburg, D-97074 W\"urzburg, Germany
\and University of Geneva, Chemin d'Ecogia 16, CH-1290 Versoix, Switzerland
\and Finnish MAGIC Group: Finnish Centre for Astronomy with ESO, University of Turku, FI-20014 Turku, Finland
\and Departament de F\'isica, and CERES-IEEC, Universitat Aut\`onoma de Barcelona, E-08193 Bellaterra, Spain
\and Japanese MAGIC Group: Physics Program, Graduate School of Advanced Science and Engineering, Hiroshima University, 739-8526 Hiroshima, Japan
\and Armenian MAGIC Group: ICRANet-Armenia at NAS RA, 0019 Yerevan, Armenia
\and Croatian MAGIC Group: University of Split, Faculty of Electrical Engineering, Mechanical Engineering and Naval Architecture (FESB), 21000 Split, Croatia
\and Croatian MAGIC Group: Josip Juraj Strossmayer University of Osijek, Department of Physics, 31000 Osijek, Croatia
\and Japanese MAGIC Group: Department of Physics, Kyoto University, 606-8502 Kyoto, Japan
\and Japanese MAGIC Group: Department of Physics, Tokai University, Hiratsuka, 259-1292 Kanagawa, Japan
\and Universit\`a di Siena and INFN Pisa, I-53100 Siena, Italy
\and Saha Institute of Nuclear Physics, A CI of Homi Bhabha National Institute, Kolkata 700064, West Bengal, India
\and Inst. for Nucl. Research and Nucl. Energy, Bulgarian Academy of Sciences, BG-1784 Sofia, Bulgaria
\newpage
\and Japanese MAGIC Group: Department of Physics, Yamagata University, Yamagata 990-8560, Japan
\and Finnish MAGIC Group: Space Physics and Astronomy Research Unit, University of Oulu, FI-90014 Oulu, Finland
\and Croatian MAGIC Group: Ru\dj{}er Bo\v{s}kovi\'c Institute, 10000 Zagreb, Croatia
\and Japanese MAGIC Group: Institute for Space-Earth Environmental Research and Kobayashi-Maskawa Institute for the Origin of Particles and the Universe, Nagoya University, 464-6801 Nagoya, Japan
\and INFN MAGIC Group: INFN Sezione di Perugia, I-06123 Perugia, Italy
\and INFN MAGIC Group: INFN Roma Tor Vergata, I-00133 Roma, Italy
\and Japanese MAGIC Group: Department of Physics, Konan University, Kobe, Hyogo 658-8501, Japan
\and also at International Center for Relativistic Astrophysics (ICRA), Rome, Italy
% ...
}

   \date{November 22, 2022}

  \abstract
   {The BL Lac object 1ES\,0647+250 is one of the few distant $\gamma$-ray emitting blazars detected at very high energies (VHEs; $\gtrsim$100\,GeV) during a non-flaring state. It was detected with the MAGIC telescopes during a period of low activity in the years 2009-2011 as well as during three flaring activities in the years 2014, 2019, and 2020, with the highest VHE flux in the last epoch. An extensive multi-instrument data set was collected as part of several coordinated observing campaigns over these years.}
   {We aim to characterise the long-term multi-band flux variability of 1ES\,0647+250, as well as its broadband spectral energy distribution (SED) during four distinct activity states selected in four different epochs, in order to constrain the  physical parameters of the blazar emission region under certain assumptions.}
   {We evaluated the variability and correlation of the emission in the different energy bands with the fractional variability and the Z-transformed discrete correlation function, as well as its spectral evolution in X-rays and $\gamma$ rays. Owing to the controversy in the redshift measurements of 1ES\,0647+250 reported in the literature, we also estimated its distance in an indirect manner through a comparison of the GeV and TeV spectra from simultaneous observations with \textit{Fermi}-LAT and MAGIC during the strongest flaring activity detected to date. Moreover, 
   we interpret the SEDs from the four distinct activity states within the framework of one-component and two-component leptonic models, proposing specific scenarios that are able to reproduce the available multi-instrument data. }
   {We find significant long-term variability, especially in X-rays and VHE $\gamma$ rays. Furthermore, significant (3-4$\sigma$) correlations were found between the radio, optical, and high-energy (HE) $\gamma$-ray fluxes, with the radio emission delayed by about $\sim$400 days with respect to the optical and $\gamma$-ray bands. The spectral analysis reveals a harder-when-brighter trend during the non-flaring state in the X-ray domain. {However, no clear patterns were observed for either the enhanced states or} the HE (30 MeV<E<100 GeV) and VHE $\gamma$-ray emission of the source. The indirect estimation of the redshift yielded a value of $z=0.45\pm0.05$, which is compatible with some of the values reported in the literature. The SEDs related to the low-activity state and the three flaring states of 1ES\,0647+250 can be described reasonably well with the both one-component and two-component leptonic scenarios. However, the long-term correlations indicate the need for an additional {radio-producing} region located about 3.6 pc downstream {from the gamma-ray producing region.}
   }
   {}


   \maketitle
%

\section{Introduction}
Blazars are radio-loud active galactic nuclei whose relativistic jets point towards the Earth. Blazars can be classified according to the spectral features in the optical band {as} BL Lacertae (BL Lacs) objects and flat spectrum radio quasars (FSRQs).
While BL Lacs have an (almost) featureless optical spectrum, FSRQs show strong, broad emission lines in the optical band \citep{urry1995}. Blazars are also characterised by high variability over very different timescales. They emit mostly non-thermal radiation at all wavelengths, from radio to $\gamma$ rays. However, most of the blazars detected in very-high-energy (VHE; $\gtrsim$100\,GeV) $\gamma$ rays are BL Lacs.

Blazars display a spectral energy distribution (SED) characterised by the presence of two bumps \citep{ghisellini2017}. The first bump originates from synchrotron radiation by relativistic electrons. The second peak is commonly explained by a leptonic scenario through inverse Compton (IC) scattering of synchrotron photons -- synchrotron self-Compton (SSC) -- scattering with the same electron population \citep[see e.g.][]{celotti2008,ghisellini2010} and/or IC scattering of photons coming from outside the jet in an external Compton process \citep{dermer1994}. Alternatively, different models with a hadronic origin have been proposed to explain the high-energy bump in the SED of blazars \citep[e.g.][]{mannheim1993, cerruti2015}. BL Lacs can be divided into three groups depending on the frequency of the synchrotron peak: low- ($\nu_{\rm peak}$~<~10$^{14}$\,Hz), intermediate- (10$^{14}$\,Hz~<~$\nu_{\rm peak}$~<~10$^{15}$\,Hz), and high-energy-peaked BL Lacs (HBLs; $\nu_{\rm peak}$~>~10$^{15}$\,Hz;  \citealt{padovani1995}). Another category of BL Lacs was introduced by \cite{costamante2001}, naming those whose peak is above $\nu_{\rm peak}$~>~10$^{17}$\,Hz extreme HBLs (EHBLs). These objects display a high X-ray flux with respect to their optical/UV emission. They can also show a high-energy peak shifted to VHE $\gamma$-ray frequencies \citep[see for instance the BL Lac 1ES\,0229+200 and the sources reported by][]{acciari2020b}.


1ES\,0647+250 is a BL Lac object previously catalogued as an HBL \citep{costamante2002,aleksic2011}. It has an uncertain redshift, with various values reported in the literature. A lower limit of $z > 0.6$ was first derived from imaging of the source by \cite{falomo1999}. However, based on deep observations of the host galaxy, \cite{meisner2010} derived a value of $z=0.45^{+0.11}_{-0.10}$, and \cite{kotilainen2011} estimated a redshift of $z=0.41 \pm 0.06$ after using the imaging redshift method from \cite{sbarufatti2005}. Spectral lines have not been detected in its spectrum. These non-detections were used to derive lower limits on the redshift of $z>0.3$ by \cite{scarpa2000} and $z>0.47$ by \cite{sbarufatti2005}. An accurate {measurement} of the redshift is still lacking, though the most recent work by \cite{paiano2017} set a lower limit of $z > 0.29$.


It was first reported as a VHE $\gamma$-ray emitter by the Major Atmospheric Gamma-ray Imaging Cherenkov (MAGIC) collaboration with a flux above 100~GeV of (3.0~$\pm$~0.7)\% Crab Nebula flux Units (C.U.; \citealt{delotto2012}). Later on, it was detected by the Very Energetic Radiation Imaging Telescope Array System (VERITAS) with a $\gamma$-ray flux of (2.7~$\pm$~0.7)\%~C.U. above 140~GeV as part of the VERITAS blazar programme carried out between 2010 and 2013 \citep{dumm2013}.


As for most blazars, this source is bright and variable in all the electromagnetic bands and has been observed as part of many programmes in radio \citep{piner2014}, optical \citep{kapanadze2009}, and X-rays \citep{perlman2005}. It is also detected at high energies (HEs, 30 MeV<E<100 GeV), and it can be found in each of the \textit{Fermi} Large Area {Telescope \citep[LAT;][]{2009ApJ...697.1071A}} source catalogues from 1FGL onwards, including the 4FGL catalogue \citep{abdollahi2020}.
Significant variability has been detected in the optical band, but no intra-night or short-burst variability has been claimed for this source. Furthermore, it has not been possible to come to a firm conclusion about the variability timescales of 1ES\,0647+250 in the optical band due to long gaps in the historical light curve \citep{kapanadze2009}. \cite{nilsson2018} also detected significant variability for this source using optical data from between 2002 and 2012. 


In this paper we perform the first long-term multi-wavelength (MWL) study of 1ES\,0647+250. We report on the detection in the VHE $\gamma$-ray band by MAGIC in four different epochs (2009-2011, 2014, 2019, and 2020), each of which corresponds to a different state of the source in terms of its optical, X-ray, and VHE flux. The MAGIC observations performed between 2009 and 2011 were triggered by the first studies done on \textit{Fermi}-LAT data above 10 GeV, which later on led to the first \textit{Fermi}-LAT catalogue of $>$10~GeV sources \citep[1FHL;][]{2013ApJS..209...34A}, and identified several VHE $\gamma$-ray emitter candidates with only one year of LAT data. The VHE $\gamma$-ray observations in 2014 were triggered by an optical flare detected by several optical facilities \citep{kiehlmann2014}. In December 2019, 1ES\,0647+250 showed a historically high X-ray flux \citep{kapanadze2019}, leading to the detection at VHE $\gamma$ rays of this blazar \citep{mirzoyan2019}. Finally, the source displayed its highest state in the VHE $\gamma$-ray domain in December 2020, after an X-ray activity comparable to the 2019 flare \citep{kapanadze2020}.


This paper is structured as follows: In Sect.~\ref{sec2} the data sets used in the analysis are introduced. Based on the MWL data collected in this work, in Sect.~\ref{sec3} we present MWL variability studies for the first time{ from radio to the VHE $\gamma$-ray band}. In Sect.~\ref{sec4} the spectral analysis of the X-ray and $\gamma$-ray data is performed. A redshift estimation of the source based on the $\gamma$-ray spectrum is presented in Sect.~\ref{sec5} and compared with previous measurements. In Sect.~\ref{sec6} we model for the first time the broadband SED of this source for the different observed periods and compare them to one another. In Sect.~\ref{sec7} a discussion and interpretation of the results are presented, and the main results are provided as a conclusion in Sect. \ref{sec8}.


\section{Multi-wavelength data}\label{sec2}
\subsection{VHE $\gamma$ rays: MAGIC telescopes}
MAGIC is a stereoscopic system of two 17 m imaging atmospheric Cherenkov telescopes located on the Canary island of La Palma, Spain, at an altitude of $\sim$2200\,m above sea level.
They work in an energy range between 50\,GeV and tens of TeVs, with a sensitivity above 100 GeV (300 GeV) of about 2\% (about 1\%) of
the Crab Nebula flux after 25~h of observations at zenith angle ZA<30$^{\circ}$ \citep{aleksic2016}.
These characteristics make the MAGIC telescopes very well suited for blazar observations in the VHE $\gamma$-ray range.

1ES\,0647+250 was first observed by MAGIC-I in {mono mode in 2008} \citep{aleksic2011}, with no detection of the source. However, an upper limit of the integral flux above 120\,GeV of 1.6 $\times$ 10$^{-11}$\,cm$^{-2}$s$^{-1}$ was estimated. For the present work, we use approximately 45 hours of stereoscopic good-quality data taken by MAGIC between November 2009 and December 2020: 26.7 hours after quality cuts between November 2009 and March 2011. In November 2014, the observations (2.2 hours after data quality cuts) were triggered under the target of opportunity programme following an enhanced flux in the optical and HE $\gamma$-ray (above 10~GeV) bands, which was measured with the procedure described in \cite{pacciani2018}. A historically high X-ray flux triggered the observations in December 2019 (2.7 hours after cuts) and in December 2020 (13.5 hours after cuts).
The data were analysed using the MAGIC Analysis and Reconstruction Software \citep[MARS;][]{zanin2013,aleksic2016}.

The data analysis was performed by separating the data into four different epochs: the first corresponds to the data taken from 2009 to 2011 (MJD 55131-55620, hereafter epoch E1); while the second (MJD 56986-56987, E2), third (MJD 58819-58821, E3) and fourth (MJD 59198-59208, E4) correspond to the target of opportunity observations in 2014, 2019, and 2020, respectively. Table~\ref{significances} shows the significances of the detection during the different epochs as estimated following Eq.~17 in \cite{li1983}.

\begin{table*}[]
\caption{Significance and integrated flux above 100 GeV of the different detections of 1ES\,0647+250.}
\centering
\begin{tabular}{ccccccc}
\hline
Epoch & \begin{tabular}[c]{@{}c@{}}Year(s) of \\ observation\end{tabular}  & \begin{tabular}[c]{@{}c@{}}Time interval \\ {[}MJD{]}\end{tabular} & \begin{tabular}[c]{@{}c@{}}Live time\\{[}h{]}\end{tabular} & \begin{tabular}[c]{@{}c@{}}Significance\\ {[}Li\&Ma{]}\end{tabular} & \begin{tabular}[c]{@{}c@{}} $f \ (>100 \ \text{GeV})$ \\ {[}10$^{-11}\cdot$ cm$^{-2} \cdot$ s$^{-1}${]}\end{tabular} & \begin{tabular}[c]{@{}c@{}} $f \ (>100 \ \text{GeV})$ \\ {[}\% C.U.{]}\end{tabular} \\ \hline
E1 & 2009-2011 &  55131.0 - 55620.9 & 26.7   &  5.5$\sigma$ & 0.97 $\pm$ 0.24 & 2.0 $\pm$ 0.5 \\ 
E2 & 2014      &  56986.2 - 56987.2 & 2.2   &  5.3$\sigma$ & 1.62 $\pm$ 0.78 & 3.4 $\pm$ 1.6 \\ 
E3 & 2019      &   58819.0 - 58821.2 & 2.7   &  6.1$\sigma$ & 3.82 $\pm$ 0.88 & 8.0 $\pm$ 1.8 \\ 
E4 & 2020      &  59198.0 - 59208.0 & 13.5   & 22.9$\sigma$ & 7.10 $\pm$ 0.45 & 15.0 $\pm$ 1.0 \\ \hline
\end{tabular}
\tablefoot{Live time is given after data quality cuts. }
\label{significances}
\end{table*}

\subsection{HE $\gamma$ rays: \textit{Fermi}-LAT}

The GeV $\gamma$-ray emission from 1ES\,0647+250 was characterised with the LAT on board the {\it Fermi} Gamma-ray Space Telescope.  The \textit{Fermi}-LAT data presented 
in this paper were analysed using the standard \textit{Fermi} analysis software tools 
(version \textit{v11r07p00}), and the \textit{P8R3\_SOURCE\_V2} response function. We used events from $0.3-300$\,GeV selected within a 10$^\circ$ radius region of interest (ROI) centred 
on 1ES\,0647+250 and having a zenith distance
below 100$^\circ$ to avoid contamination from the Earth's limb. The usage of events above 0.3\,GeV (instead of above 0.1\,GeV) is advantageous for sources with hard $\gamma$-ray spectra (photon index $<$2.0), especially if the source is weak and long integration times are needed for significant detections. The higher minimum energy somewhat reduces the detected number of photons from the source, but this effect is small for hard sources. On the other hand, the angular resolution (68\% containment) improves from $\sim$5$^\circ$ to $\sim$2$^\circ$ when increasing the energy from 0.1\,GeV to 0.3\,GeV, which reduces the diffuse backgrounds (which are always softer than photon index 2), {thus making} the analysis less sensitive to possible contamination from non-accounted (transient) neighbouring sources, and {reducing} the systematic uncertainties. The diffuse Galactic and isotropic components were modelled with the files  gll\_iem\_v07.fits and
iso\_P8R3\_SOURCE\_V2\_v1.txt, 
respectively\footnote{\url{https://fermi.gsfc.nasa.gov/ssc/data/access/lat/BackgroundModels.html}}. All point sources in the fourth {\it Fermi}-LAT source catalogue \citep[4FGL,][]{abdollahi2020} located in the 10$^\circ$ ROI and an additional surrounding 5$^\circ$ wide annulus were included in the model. In the unbinned likelihood fit, the normalisation and spectral parameters of all the sources were fixed to the 4FGL values, with the exception of the seven sources within the ROI identified as variable and with a detection significance larger than 10~$\sigma$, where the normalisation parameters were allowed to vary. For the three objects located within an angular distance of 5$^\circ$ of 1ES\,0647+250 (i.e. 4FGL~J0650.6+2055, 4FGL~J0653.7+2815, and 4FGL~J0709.1+2241), the spectral parameters were also allowed to vary. The normalisation of the diffuse components (Galactic and isotropic) was also allowed to vary in the unbinned likelihood fits. In the 4FGL-DR3 \citep{abdollahi2020,abdollahi2022}, which integrates over 12 years, the log-parabola (LogP) function is preferred to reproduce the spectrum over the power-law (PL) function with a significance of 4.1$\sigma$. However, owing to the much shorter timescales used in this study (shorter than 2 years), we decided to parameterise the $\gamma$-ray spectral shape of  1ES\,0647+250 with a PL, where both the normalisation (flux) and the PL index were kept as free parameters.

Owing to the moderate sensitivity of \textit{Fermi}-LAT for the detection of 1ES\,0647+250 on day/week timescales (especially when the source is not flaring), we performed the unbinned likelihood analysis on consecutive 30-day time intervals (not centred on the MAGIC time) to determine the light curve in the energy band $0.3-300$\,GeV, as reported in  Fig.~\ref{1ES0647LCs}. The source is detected with a maximum-likelihood test statistic (TS)\footnote{The maximum-likelihood TS \citep{1996ApJ...461..396M} is defined as TS = 2$\Delta$ log(likelihood) between models with and without a point source at the position of 1ES\,0647+250.} above 4 for most of the 30-day time intervals. There are seven 30-day time intervals that yielded a TS below 4, for which we computed 95\% confidence level upper limits using a fixed PL index of 1.70, {reported in the 4FGL-DR3 catalogue \citep{abdollahi2022}}. The PL spectral index of the source for each time bin is computed using the same procedure as the light curve analysis.

For the flaring episodes in 2014, 2019, and 2020, where we wanted to combine the data with VHE $\gamma$-ray spectra from MAGIC, we decided to use a time interval of 12 days for the first two, and 10 days for the third. We also computed the spectrum contemporaneous to the 2-year-long MWL observations in 2009-2011. The spectral results are reported in Sects.~\ref{sec4} and \ref{sec6}.

\subsection{X-ray observations: \textit{Swift}-XRT}
The X-Ray Telescope \citep[XRT;][]{burrows2004} on the \textit{Neil Gehrels Swift} Observatory carried out 70 distinct observations of this blazar between May 2010 and December 2020. In particular, the source was observed several times distributed in the different observing campaigns previously defined. \textit{Swift}-XRT pointed to 1ES~0647+250 a total of 25 times
during E1, between May 2010 and March 2011 (MJD 55322-55623). Moreover, the source was also observed during the different flaring states in E2, E3 and E4. It was observed five times in November 2014 (E2, MJD 56981-56987). Another 19 observations were performed during and after the enhanced activity of 2019 (E3), from December 2019 until March 2020 (MJD 58816-58914). Finally, it was targeted eight more times in December 2020 (E4, MJD 59196-59207). An additional 13 more observations were performed, non-simultaneously to those from MAGIC. {The \textit{Swift}-XRT observations were carried out in the windowed timing and photon counting readout modes. The data were processed using the XRTDAS software package (v.3.6.0), which was developed by the Agenzia Spaziale Italiana (ASI) Space Science Data Center (SSDC) and released by HEASARC in the HEASoft package (v.6.28). The data were calibrated and cleaned with standard filtering criteria using the \texttt{xrtpipeline} task and the calibration files available from the \textit{Swift}-XRT CALDB (version 20200724).} {For the spectral analysis, events were selected within a circle of 20-pixel ($\sim$46 arcsecond) radius, which encloses about 90\% of the point spread function, centred at the source position. The background was estimated from nearby circular regions with a radius of 40 pixels.} {The ancillary response files were generated using the \texttt{xrtmkarf} task applying corrections for point spread function losses and Charge-Coupled Device (CCD) defects using the cumulative exposure maps. Before the spectral fitting, the 0.3–10~keV source energy spectra were binned using the \texttt{grppha} FTOOL to ensure a minimum of 20 counts per bin.}


\subsection{UV/optical observations: \textit{Swift}-UVOT}
The Ultra-Violet and Optical Telescope \citep[UVOT;][]{roming2005} on board the \textit{Swift} satellite, \textit{Swift}-UVOT, has performed photometric observations in three optical (U, B and V) and three UV (UVW1, UVM2, and UVW2) filters, for a total number of 70 {observations} from May 2010 to December 2020. All the UVOT observations are simultaneous to those performed by XRT.

We evaluated aperture photometry for each total exposure applying the official software included in HEASoft package (v6.23), with a final check for attitude stability. We extracted the source counts within the standard circular aperture of 5$\arcsec$ radius, and the background counts from an annular region of inner radius 26$\arcsec$ and 9$\arcsec$ size. We applied the official calibrations \citep{2008MNRAS.383..627P,Brev2011} from the \textit{Swift}-UVOT CALDB (version 20201026) to convert source counts to fluxes and then, a mean Galactic $E(B-V)$ value of 0.0835 mag \citep{Schl2011} and an interstellar extinction curve \citep{Fitz1999} were used to obtain $\nu F(\nu)$ values at filter effective frequencies.

\subsection{Optical data}
Optical monitoring of the source in the R band was also performed by the Tuorla blazar monitoring programme\footnote{\url{http://users.utu.fi/kani/1m/}}. For these observations, the 35 cm Kungliga Vetenskapsakademien (KVA) telescope, located in La Palma, was used. The data analysis was performed following the procedure described in \cite{nilsson2018}. This analysis includes the subtraction of the stellar emission from the host galaxy and the correction for Galactic extinction. The monitoring of this source started in December 2002, and continued until December 2019 (MJD 52615-58835).

Optical observations of 1ES\,0647+250 were also performed in December 2020 with the 0.4 m robotic telescopes of Las Cumbres Observatory \citep[LCOGT;][]{brown2013}. During this period, the source was also observed by the robotic 2.0 m Liverpool Telescope (LT) at the Roque de los Muchachos Observatory in La Palma \citep{steele2004}. These observations were performed with the Infrared-Optical (IO) instrument and its optical imaging component, the IO:O. Furthermore, it was observed during the night of 22 December 2020 by the 43 cm PIRATE (Physics Innovation Robotic Astronomical Telescope Explorer) telescope located at the Teide Observatory, on the Canary island of Tenerife \citep{holmes2011}. 


\subsection{Radio observations: OVRO}
1ES\,0647+250 is also part of the Owens Valley Radio Observatory (OVRO) blazar monitoring programme \citep{richards2011}\footnote{\url{https://www.astro.caltech.edu/ovroblazars/}}. These observations were conducted with the OVRO 40 m radio telescope, working at a frequency of 15\,GHz. The source was monitored by OVRO from January 2008 until December 2020, covering all the MAGIC observing periods (MJD 54476-59199). The data reduction was performed according to the procedure described in \cite{richards2011}. Observations with a signal-to-noise ratio $<3$ were treated as non-detections and thus were not included in the MWL light curve and analysis. This resulted in 446 observations after excluding these measurements from the analysis.

\section{Multi-wavelength light curve analysis}\label{sec3}
\begin{figure*}
        \includegraphics[width=\textwidth]{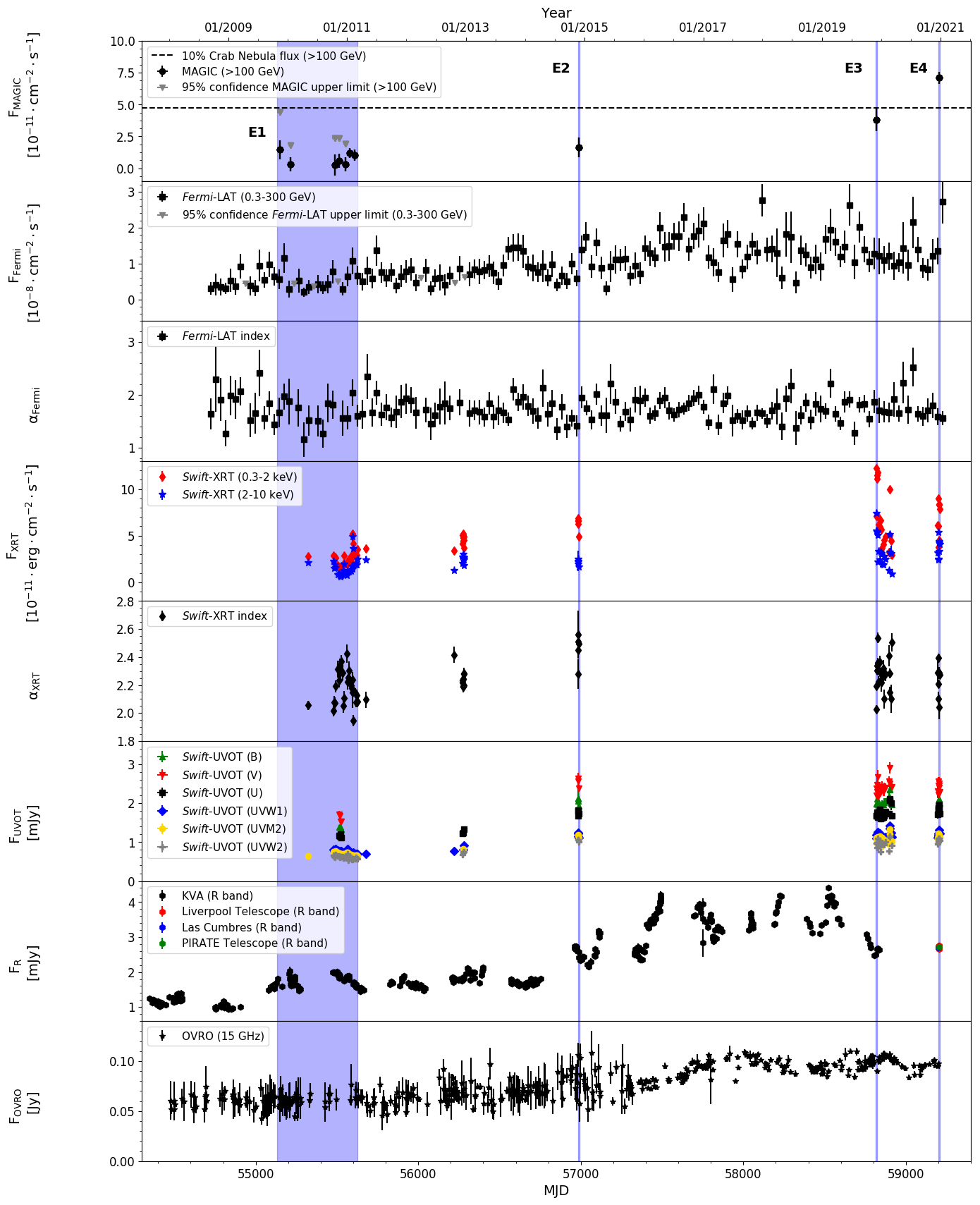}
    \caption{MWL light curves of 1ES\,0647+250. From top to bottom: MAGIC (>100\,GeV, 30-day binning), \textit{Fermi}-LAT flux and spectral index (300\,MeV-300\,GeV, 30-day binning), \textit{Swift}-XRT (0.3-2\,keV and 2-10\,keV), \textit{Swift}-XRT spectral index, \textit{Swift}-UVOT (B, V, U, UVW1, UVM2, and UVW2 filters), optical R band (KVA, Liverpool Telescope, Las Cumbres, and PIRATE telescope), and OVRO (15\,GHz). We note that all the optical observations except those from 2020 were performed by KVA. The smaller error bars of the OVRO light curve after 2016 are due to a major upgrade of the instrument. Blue contours correspond to the MAGIC campaigns during the different observed states. }
    \label{1ES0647LCs}
\end{figure*}

The MWL light curves of 1ES\,0647+250, from VHE $\gamma$ rays to radio wavelengths, are presented in Fig.~\ref{1ES0647LCs}. All the curves are daily binned except for the MAGIC and \textit{Fermi}-LAT light curves, for which 30-day binning is used due to the limited ability of these two instruments to detect 1ES\,0647+250 in the HE and VHE $\gamma$-ray bands when the source is not flaring. In the following subsections, the variability and interband correlations of the MWL data set are evaluated. A description of the light curves, with maximum, mean and minimum flux values for each band, is included in the Appendix \ref{appendix}.

\subsection{Variability}



{Emission from blazars is known to be variable across the electromagnetic spectrum. We performed a variability analysis, that is, we searched for significant flux variations and patterns in the data at different timescales on the light curves presented in Fig.~\ref{1ES0647LCs} by testing the steady-flux hypothesis in the different bands.}

The sparse time coverage and the large number of times that the source was not detected significantly with MAGIC prevent one from performing a reliable variability analysis in the VHE $\gamma$-ray energy range. In the HE band, this source is catalogued as variable {in the 4FGL-DR3 \textit{Fermi}-LAT catalogue} \citep{abdollahi2020,abdollahi2022}, with a variability index of $\sim$315, estimated as defined in Table 6 of \cite{abdollahi2022}. A fit to a constant function to the \textit{Fermi}-LAT fluxes reported in Fig.~\ref{1ES0647LCs} {yields a $\chi^2/\text{d.o.f.}=346/143 \simeq 2.4$} ($p_{value} = 10^{-20}$). Therefore, the HE emission from this source is clearly variable on timescales of 30 days, showing an increasing trend in the flux over time. This long-term trend was evaluated by a linear fit with an increasing flux, {resulting in a $\chi^2/\text{d.o.f.}=191/142 \simeq 1.3$} ($p_{value} = 0.004$), which is preferred to a fit to a constant average flux. Due to the low HE flux level and limited sensitivity of the LAT, we cannot detect possible variations in timescales shorter than one month.



In radio to X-rays the source is significantly variable on long-term timescales {of} the order of several months or years, as is typical for blazars and in line with previous studies for this source \citep[see e.g.][]{kapanadze2009,kiehlmann2014,nilsson2018,kapanadze2019,kapanadze2020}. The results of the variability analysis and the goodness of the constant fit performed for each wave band are shown in Table \ref{variability}. Moreover, the optical R-band and radio light curves also show the same increasing trend observed in the HE $\gamma$-ray light curve. An increasing linear fit is able to describe this steady flux increase over the years. However, since these bands also show variability {on} shorter timescales, as observed in Fig. \ref{1ES0647LCs}, the $\chi^2/\text{d.o.f.}$ of both fits is still >1. Additionally, we also investigated the variability of the data from E1 and E4, as they are the periods with best MWL coverage. The results are also displayed in Table \ref{variability}. Significant variability in the X-ray band was found during these two epochs, as well as variability in the optical R-band for E1. 


\begin{table}[]
\caption{Goodness of the constant flux hypothesis for every MWL light curve.} 
\centering
\begin{adjustbox}{max width=\columnwidth}
\begin{tabular}{cccc}
\hline
\multirow{2}{*}{Waveband} & \multicolumn{3}{c}{$\chi^{2}/$d.o.f.} \\ \cline{2-4}
                    &  2009-2020                 &  2009-2011 (E1)        & 2020 (E4) \\ \hline
HE $\gamma$ rays    &  346/143 $\simeq$ 2.4      &  15.7/14 $\simeq$ 1.1  &   --   \\ 
X-rays (2-10 keV)   &  3255/69 $\simeq$ 47.2     &  582/24 $\simeq$ 24.3  &   154/7 = 22.0         \\
X-rays (0.3-2 keV)  &  21575/69 $\simeq$ 312.7   &  1159/24 $\simeq$ 48.3 &   995/7 $\simeq$ 142.1 \\
UV (uvw2 filter)    &  1996/55 $\simeq$ 36.3     &  66.6/23 $\simeq$ 2.9  &   6.3/6 $\simeq$ 1.1   \\ 
Optical B band      &  338/31 $\simeq$ 10.9      &  1.3/3 $\simeq$ 0.4    &   4.5/7 $\simeq$ 0.6   \\ 
Optical R band      &  129851/550 $\simeq$ 236.1 &  922/76 $\simeq$ 12.1  &   1.1/12 $\simeq$ 0.1  \\ 
Radio (15 GHz)      &  7169/446 $\simeq$ 16.1    &  26/50 $\simeq$ 0.5    &   --   \\ \hline

\end{tabular}
\end{adjustbox}
\label{variability}
\end{table}

\begin{figure}
        \includegraphics[width=\columnwidth]{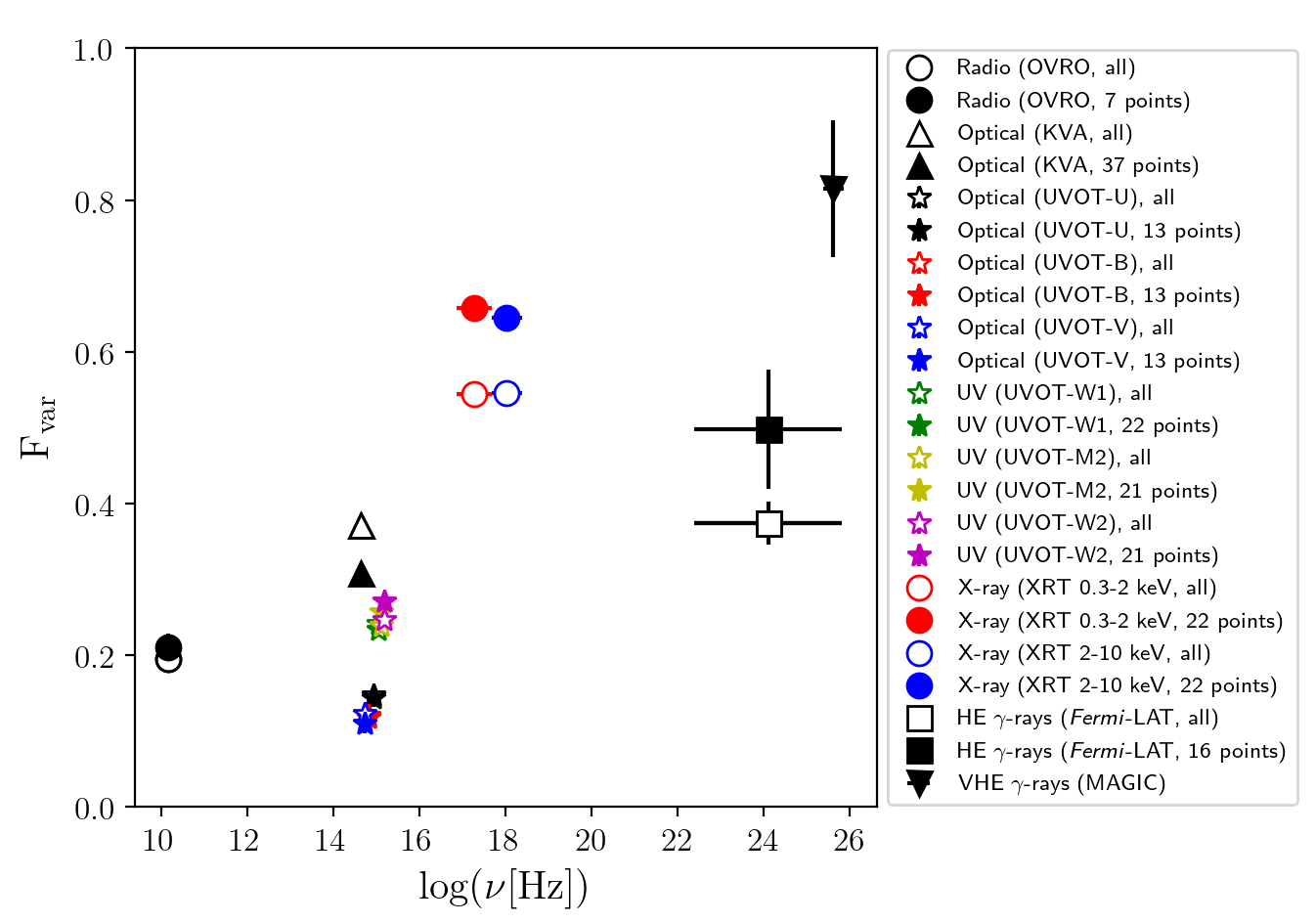}
    \caption{Fractional variability of 1ES\,0647+250. Filled markers represent MWL observations quasi-simultaneous to the MAGIC observations. Open markers correspond to the $F_{var}$ of the complete data sets. {Filled markers represent the $F_{var}$ using only the simultaneous MWL data with respect to the MAGIC observations}.}
    \label{1ES0647_fractional_variability}
\end{figure}

{Moreover, it is also important to quantify the amount of variability displayed in each band. This can provide useful information about the dynamics of the particle population that dominates the emission in the energy band probed. For this purpose, we used} 
the fractional variability as a measurement of the degree of variability. We followed the prescription of \cite{vaughan2003}, where this parameter is estimated as \begin{equation}
F_{var}=\sqrt{\frac{S^2-\langle\sigma^2_{err}\rangle}{\langle x \rangle^2}},
\label{fractional_variability_equation}
\end{equation}
where $\langle x \rangle$ and $S^2$ are the mean and the variance of the distribution of measured fluxes, respectively, and $\langle \sigma^2_{err} \rangle$ mean square error of the data.
The uncertainty associated with the fractional variability is estimated following the prescription of \cite{poutanen2008}, as described by \cite{aleksic2015b}:

\begin{equation}
\Delta F_{var}=\sqrt{F^2_{var}-err(\sigma^2_{NXS})}-F_{var},
\label{fractional_variability_error_equation}
\end{equation}

\noindent where $err(\sigma^2_{NXS})$ is the normalised excess variance taken from \cite{vaughan2003}, calculated as

\begin{equation}
err(\sigma^2_{NXS})=\sqrt{\left( \sqrt{\frac{2}{N}}\frac{\langle\sigma^2_{err}\rangle}{\langle x \rangle^2} \right)^2 + \left( \sqrt{\frac{\langle\sigma^2_{err}\rangle}{N}} \frac{2F_{var}}{\langle x \rangle} \right)^2},
\label{err_sigma_nxs_equation}
\end{equation}

\noindent where $N$ is the number of data points. A deeper discussion on the estimation and caveats of the fractional variability is presented in \cite{aleksic2015b} and \cite{schleicher2019}, and references therein. The results of the fractional variability for 1ES\,0647+250 are presented in Fig.~\ref{1ES0647_fractional_variability}. 

Owing to the remarkably different temporal coverage of 1ES~0647+250 at 15 GHz, R band, and HE $\gamma$ rays (where the data span over the entire multi-year period considered in this study), in comparison to the UV, X-rays and VHE $\gamma$ rays, we decided to apply two strategies to quantify the fractional variability. On the one hand, we used all flux values reported in the {light curves} from Fig.~\ref{1ES0647LCs} to compute the variability at radio, R-band, UV, X-ray, and HE $\gamma$-ray energies. The results are displayed with open markers in Fig.~\ref{1ES0647_fractional_variability}, and show a slight increase in the overall flux variability with increasing energy. Additionally, we computed the fractional variability for all the energy bands sampled, but this time selecting only flux measurements that related to observations that were performed quasi-simultaneous to those from MAGIC ($\pm$0.5 days). In the case of \textit{Fermi}-LAT, where the flux measurements relate to 30-day bins, we use the GeV flux from the 30-day bin that contains the MAGIC observations. The results obtained with this strategy are displayed with filled markers in Fig.~\ref{1ES0647_fractional_variability}.
The highest variability occurs at VHE $\gamma$-ray energies, although the statistical errors are large due to the relatively large flux measurement errors and the somewhat limited data set collected with MAGIC, biased towards bright flares. On the other hand, the variability at X-rays has small uncertainties, because of the smaller flux measurement errors and the larger data set, and it is clearly larger than the variability at radio, optical/UV, and HE $\gamma$ rays. Finally, for comparison, we also estimated the fractional variability of the radio/optical/UV/X-ray data sets with a 30-day binning matching the one from the HE $\gamma$-ray data, with no significant differences with respect to the values shown in Fig. \ref{1ES0647_fractional_variability}. This indicates that the short-scale variations are {less important} and the dominant variability corresponds to the long-term variability.

1ES\,0647+250 shows lower variability in radio and optical wavelengths than that at X-rays and HEs. Its fractional variability has a {double-maximum} shape {like,} for instance, Mrk 421 \citep{aleksic2015b,balokovic2016,2021A&A...655A..89M}, MAGIC~2001+439 \citep{aleksic2014} and 1ES~1959+650 \citep{kapanadze2018}. The variability increases from its minimum at radio {and optical} frequencies, reaching a maximum at X-ray wavelengths, followed by a drop at HE $\gamma$ rays and an increase in the VHE $\gamma$-ray regime.
This behaviour differs from other sources such as Mrk~501 \citep[][]{ahnen2017,ahnen2018} or TXS~0506+056 \citep{magic2022}, whose fractional variability progressively increases with the frequency, displaying its maximum variability in the VHE $\gamma$-ray domain. This may be {a} sign of different particle populations, environments and/or processes in the jet; and a higher synchrotron dominance in the jet \citep{aleksic2015b}. We note, however, that for 1ES\,0647+250, the $F_{var}$ may be biased in $\gamma$-ray energies due to the 30-day binning of the \textit{Fermi}-LAT and MAGIC data. Also, the $\gamma$-ray $F_{var}$ value shows very large statistical uncertainties due to the poor sampling and hence, it is not conclusive. Moreover, the structure of the $F_{var}$ plot may change with time, indicating that the population or the processes in the jet may also change \citep[see e.g.][where a {double-maximum} structure is also reported for Mrk 501]{furniss2015}. The discrepancy between the R-band observations and the UV and optical \textit{Swift} filters is well understood and caused by the lower coverage of the filters. {The effect of this coverage difference is especially noticeable between the optical and UV \textit{Swift} filters, causing the minimum $F_{var}$ to occur in the former band. However, when calculating the $F_{var}$ using only simultaneous optical and UV observations, one obtains the same $F_{var}$ value (within statistical uncertainties). Therefore, this indicates that the low $F_{var}$ in the optical filters is mostly due to the limited time coverage, and that, for equal time coverage, the $F_{var}$ is slightly larger at optical and UV than in the radio band.}

\subsection{Correlation}
We carried out a correlation analysis between the light curves of the different wave bands available. For this, we made use of the Z-transform discrete correlation function \citep[ZDCF;][]{alexander2013}. This tool is a modification of the classical discrete correlation function \citep[DCF;][]{edelson1988} with better performance under uneven sampling conditions. To estimate the significance of the cross-correlations, we followed the procedure described in detail in \cite{max-moerbeck2014b}. We simulated 3$\times$10$^{5}$ artificial light curves with the same sampling and power spectral density as our data set, as described in \cite{emmanoulopoulos2013}\footnote{The light curve simulation procedure implemented with the \textsc{python} package \texttt{DELCgen}, developed by \cite{connolly2015} following the prescription from \cite{emmanoulopoulos2013}, was used.}. The power spectral density slopes derived for each wave band are $\alpha_{\text{radio}}=1.26\pm0.18$, $\alpha_{R}=1.62\pm0.13$ and $\alpha_{\gamma \ \text{rays}}=0.75\pm0.41$, assuming a PL shape. This method has been widely used in the past to compute the significance of auto-correlation and cross-correlation studies \citep[see e.g.][]{max-moerbeck2014b,lindfors2016,otero2020}. To perform this analysis, we used the radio, optical and HE $\gamma$-ray light curves from 2008 to 2019 to avoid introducing the gap present in the optical light curve between 2019 and 2020. We performed the correlation analysis for each pair of light curves.

\begin{figure*}
\centering
\subfigure{\includegraphics[width=0.67\columnwidth]{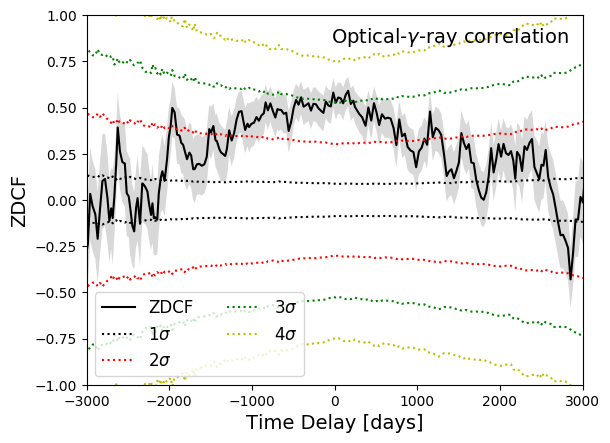}}
\subfigure{\includegraphics[width=0.67\columnwidth]{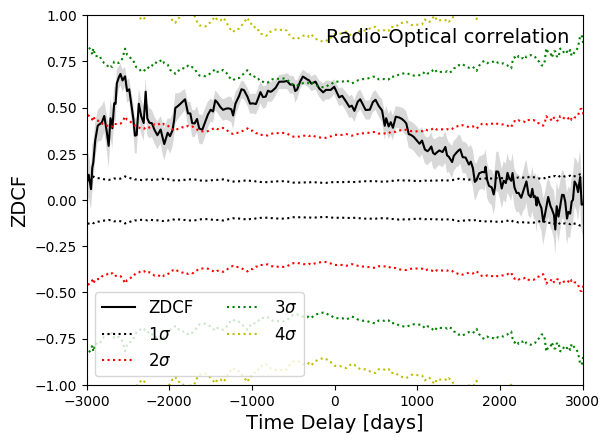}}
\subfigure{\includegraphics[width=0.67\columnwidth]{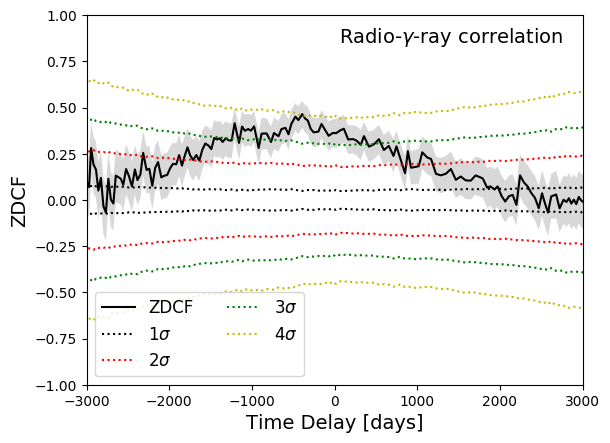}}
\caption{Long-term cross-correlation curves. The ZDCF is represented in black and its 1$\sigma$ uncertainty by the grey contour. Coloured dotted lines correspond to different significance levels, from 1$\sigma$ to 4$\sigma$. \textit{Left:} Cross-correlation between the optical and HE $\gamma$-ray light curves. \textit{Middle:} Cross-correlation between the radio and optical light curves. \textit{Right:} Cross-correlation between the radio and HE $\gamma$-ray light curves.}
\label{correlations}
\end{figure*}

{The results of the correlation analysis are shown in Fig. \ref{correlations}. We found a maximum positive correlation between the optical R-band and the HE $\gamma$-ray light curves of r=0.60 with a significance of $\sim$3$\sigma$ with respect to the no-correlation hypothesis, for a time lag of -17~days. Moreover, a long-term correlation (r=0.67, $\sim$3$\sigma$ significance) was also found between the radio and optical light curves, with its maximum degree of correlation observed at a delay of -398~days. Finally, a 4$\sigma$ long-term correlation (r=0.50) was also observed between the radio and HE $\gamma$-ray wavelengths, with the maximum correlation found at a similar delay as the radio-optical pair, -393~days, meaning that the radio emission is delayed with respect to the optical and $\gamma$-ray bands. However, the relatively wide range of time lags for which the correlation remains highly significant (which includes a 3$\sigma$ correlation at time lag zero for the light curves at radio and $\gamma$ rays), indicate that the correlation is dominated by the long-term trend, and excludes that the correlation occurs only for a specific time lag. In any case, one can estimate the most representative time lag and its related uncertainty using various strategies. When using the formalism described in \cite{alexander2013}, one obtains that largest ZDCF (ZDCF$_{max}$) between the optical R-band and the HE $\gamma$-ray light curves occurs at a time lag of $17 \pm 30$~days; between the radio and optical light curves, the maximum degree of correlation occurs at a delay of $-398 \pm 80$~days; and between the radio and HE $\gamma$-ray wavelengths at a delay of $-393 \pm 40$~days. Additionally, we also used the model-independent Monte Carlo flux randomisation and random subset selection method described in \cite{peterson1998} and \cite{peterson2004}, obtaining that the centroid of the DCF for correlations above 2$\sigma$ (DCF$_{cen}$) and the related 68\% confident limit uncertainties for the optical and $\gamma$-ray light curves is $-7 \pm 105$~days; for the radio and optical light curves $-380 \pm 88$~days; and for the radio and $\gamma$-ray light curves $-332 \pm 144$~days. The 95\% confidence limit uncertainties \citep[using][]{peterson2004} are $\pm$174~days and $\pm$219~days for the last two cases. Therefore, we can confirm that the highest degree of correlation between radio and the optical and $\gamma$-ray light curves occurs with a time lag: the radio emission is lagging the optical and the $\gamma$-ray emission.}
We note that for light curves with a large time coverage and these dominant long-term trends, the effects of short-term correlations are generally masked by the variations in long timescales \citep{smith1993,lindfors2016,raiteri2021}. Thus, {these} correlations refer to the aforementioned long-term trend that can be seen in the MWL light curves, where the flux increases over the years in radio, HE $\gamma$ rays, and especially in optical.

We also investigate whether the light curves showed any correlation at shorter timescales of the order of weeks or months. For this purpose, we performed a detrending of the long-term flux increase for all three of the bands. We followed the procedure described in \cite{lindfors2016} and \cite{acciari2020}, detrending the light curves in pairs (radio-optical, radio-HE $\gamma$ rays, and optical-HE $\gamma$ rays).

First, we fitted the lower-frequency light curve of each pair (e.g.\ radio light curve for the radio-optical pair) to a polynomial function. The polynomial order was determined by adding orders until the fit that minimises the $\chi^2/\text{d.o.f.}$ value was found. This function describes the long-term variation of the light curve. 

The polynomial fit was then scaled so that its variance equalled that of the high-frequency light curve, and the average flux of this curve was added. Then, the polynomial was multiplied by 0.1, 0.2, ... , 1.0 and subtracted from the high-frequency light curve. The best subtraction was determined by calculating the factor that minimises the root mean squared (RMS)\footnote{The RMS is estimated as $\text{RMS}=\sqrt{\sum(x_i-x_{mean})^2/N}$ where $x_{mean}$ is the mean flux of the light curve.} of the subtracted light curve. The result of this subtraction is a light curve where the common long-term variation of both the low and high frequency data sets is removed.

Third, the fractional contribution of the subtracted long-term slowly varying component was estimated by dividing the RMS of the original data set and the RMS of the light curve obtained after subtracting the polynomial function:
    
    \begin{equation}
    \text{Fraction}=1-\text{RMS}_{\text{subtracted light curve}}/\text{RMS}_{\text{original light curve}}
    .\end{equation}

\begin{figure*}
\centering
\subfigure{\includegraphics[width=0.67\columnwidth]{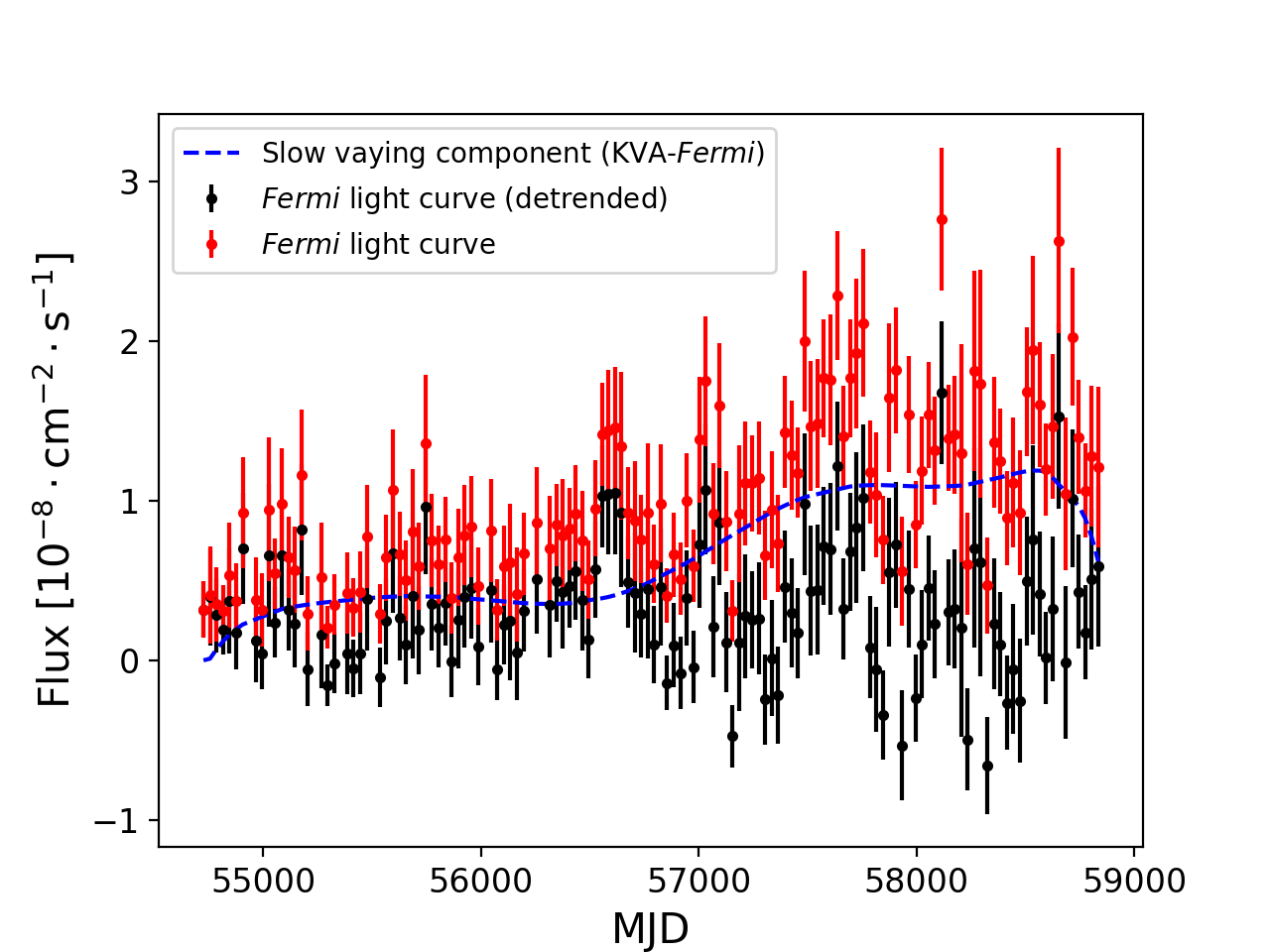}}
\subfigure{\includegraphics[width=0.67\columnwidth]{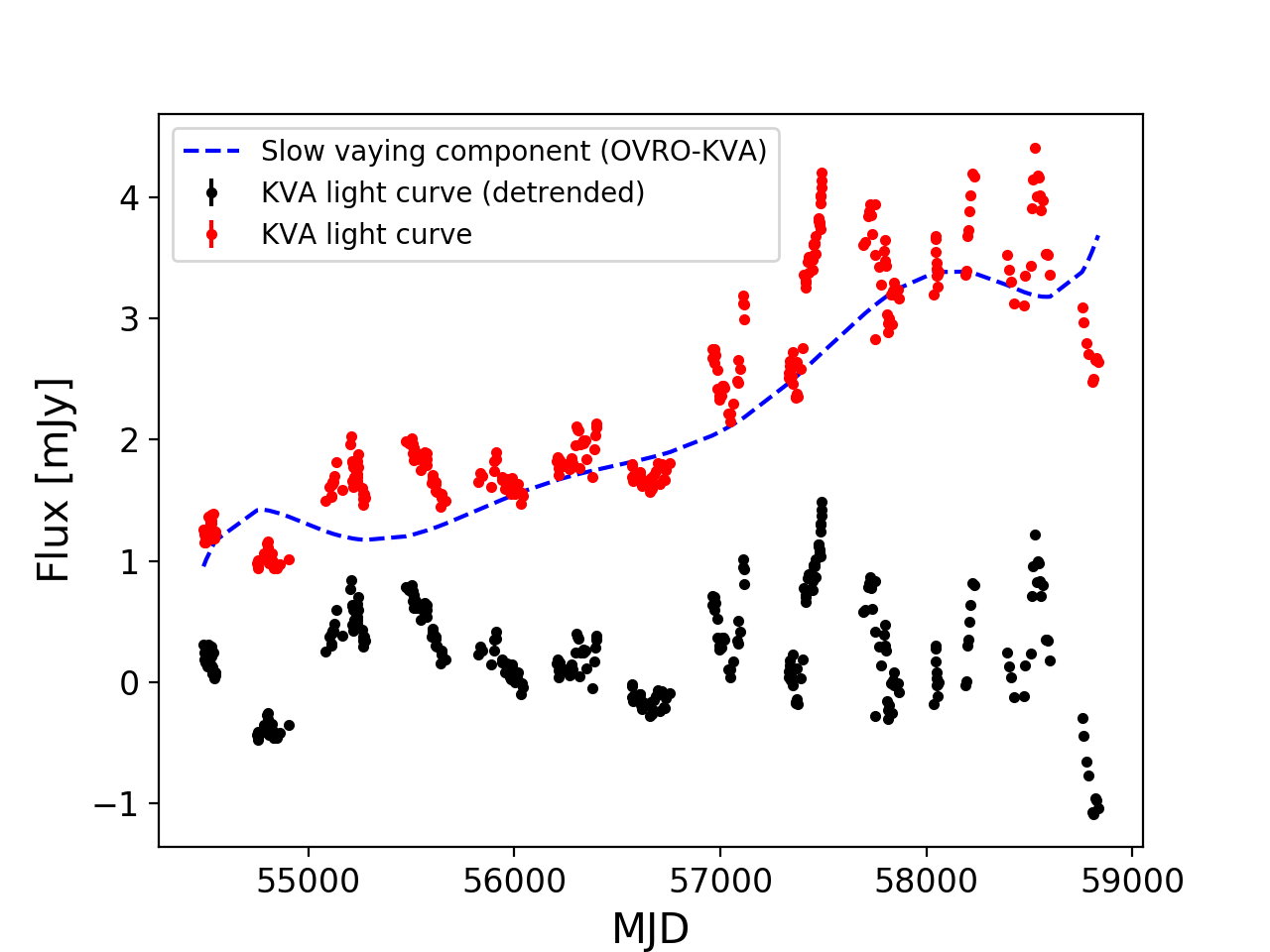}}
\subfigure{\includegraphics[width=0.67\columnwidth]{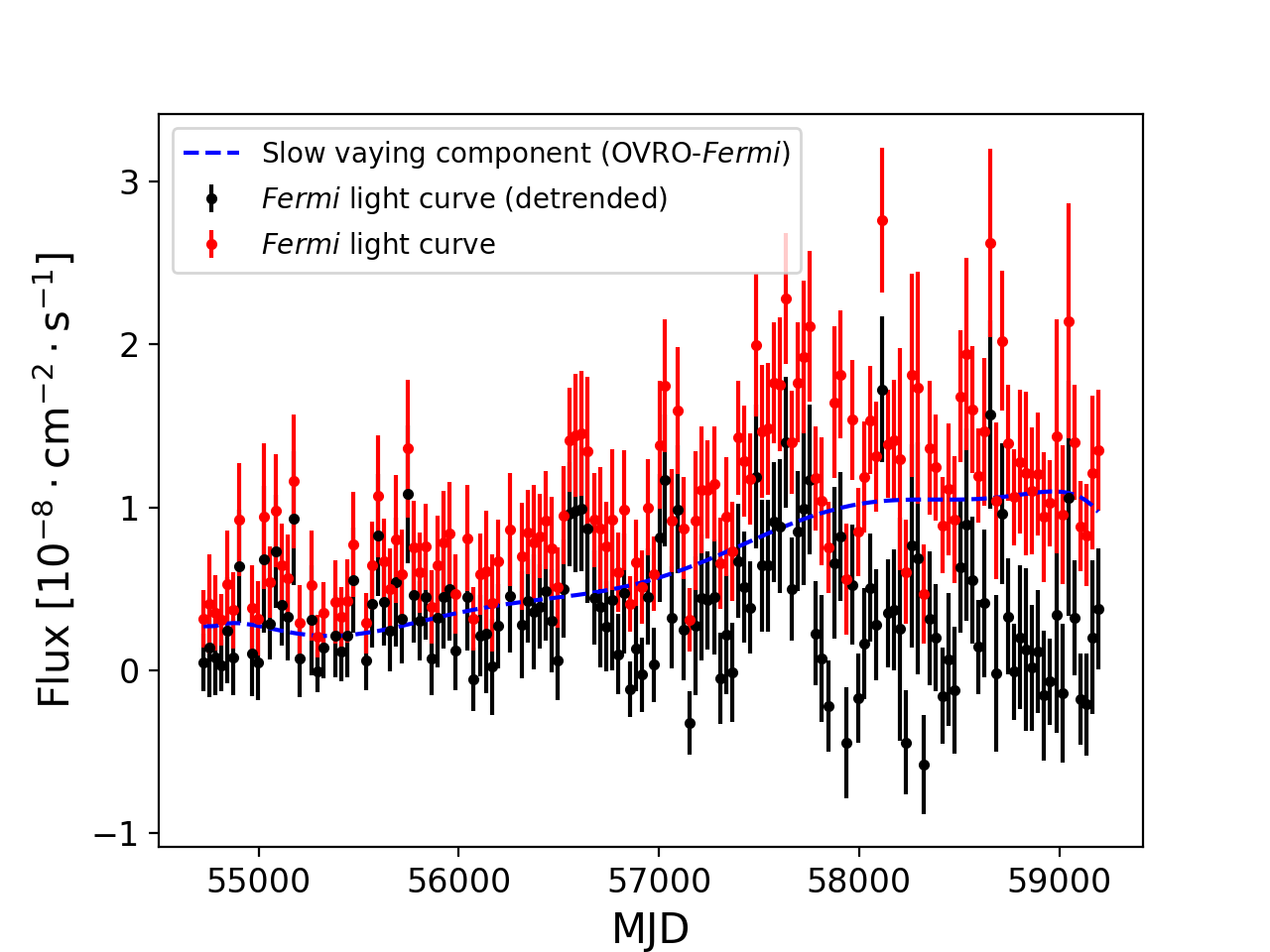}}
\caption{Light curve detrending results. Blue dashed lines represent the scaled low-frequency light curve best fit, red dots correspond to the original data set and black dots show the detrended light curve. \textit{Left:} \textit{Fermi}-LAT light curve after optical trend subtraction. \textit{Middle:} Optical light curve after radio trend subtraction. \textit{Right:} \textit{Fermi}-LAT light curve after radio trend subtraction.}
\label{light_curve_detrending}
\end{figure*}

This procedure quantifies the common slowly variable component between two wave bands. The detrended light curves can be seen in Fig.~\ref{light_curve_detrending}. Moreover, the fitted trends for the radio and optical data sets are included in the Appendix \ref{appendixb} (see Fig. \ref{detrending_fits}). We find that the slow varying component between the radio and optical light curves accounts for a fraction of 0.53 of the total variability. This value is much higher than the one reported by \cite{lindfors2016} for this source. They found that common radio-optical component has a contribution of 0.1. However, the optical light curve used for their analysis corresponds only to the first half of the data set presented in this work. We estimated the fractional contribution for both halves of our optical curve, reproducing the result presented in \cite{lindfors2016} for the data between 2008 and 2013, while we obtain a value of 0.4 for the second half of the data. This result explains the different values reported by both analyses. As for the OVRO-LAT light curve pair, this component explains a fraction of 0.23 of the flux variation. Finally, the slow component for the optical and \textit{Fermi}-LAT light curve pair is responsible for a fraction of 0.24 of the total variability. 

There may still be correlated emission on shorter timescales. In order to search for these short-term correlations, we applied the ZDCF to the detrended light curves. We do not find any correlated emission in these short timescales between radio, optical and HE $\gamma$-ray bands, with correlation coefficients <0.2. In the case of the HE $\gamma$-ray band, we do not detect significant long-term variability after detrending, {with a $\chi^2/\text{d.o.f.}=177/143 \simeq 1.2$} ($p_{value} = 0.028$). Moreover, the low coverage of the MWL data and large time bins of the \textit{Fermi}-LAT light curve due to the low flux of the source do not allow us to perform a detailed short-term correlation analysis of each individual observing epoch.

\section{Spectral analysis}\label{sec4}
\subsection{X-ray spectral analysis}\label{sec4.1}
An analysis of the X-ray spectra collected by \textit{Swift}-XRT was carried out. We compared the X-ray spectral behaviour observed in different time intervals during which the target was also observed by MAGIC. Table~\ref{table1} displays the spectral parameters and flux states during the different periods. We note that during the low state in epoch E1 and the flare from E4, the source showed significant variability in the X-ray spectrum. To account for this variability, we report the spectral parameters for the maximum, mean and minimum flux states. For the enhanced states from E2 and E3, the closest spectra in time to the MAGIC detections are shown.

\begin{figure*}
\centering
\subfigure{\includegraphics[width=0.95\columnwidth]{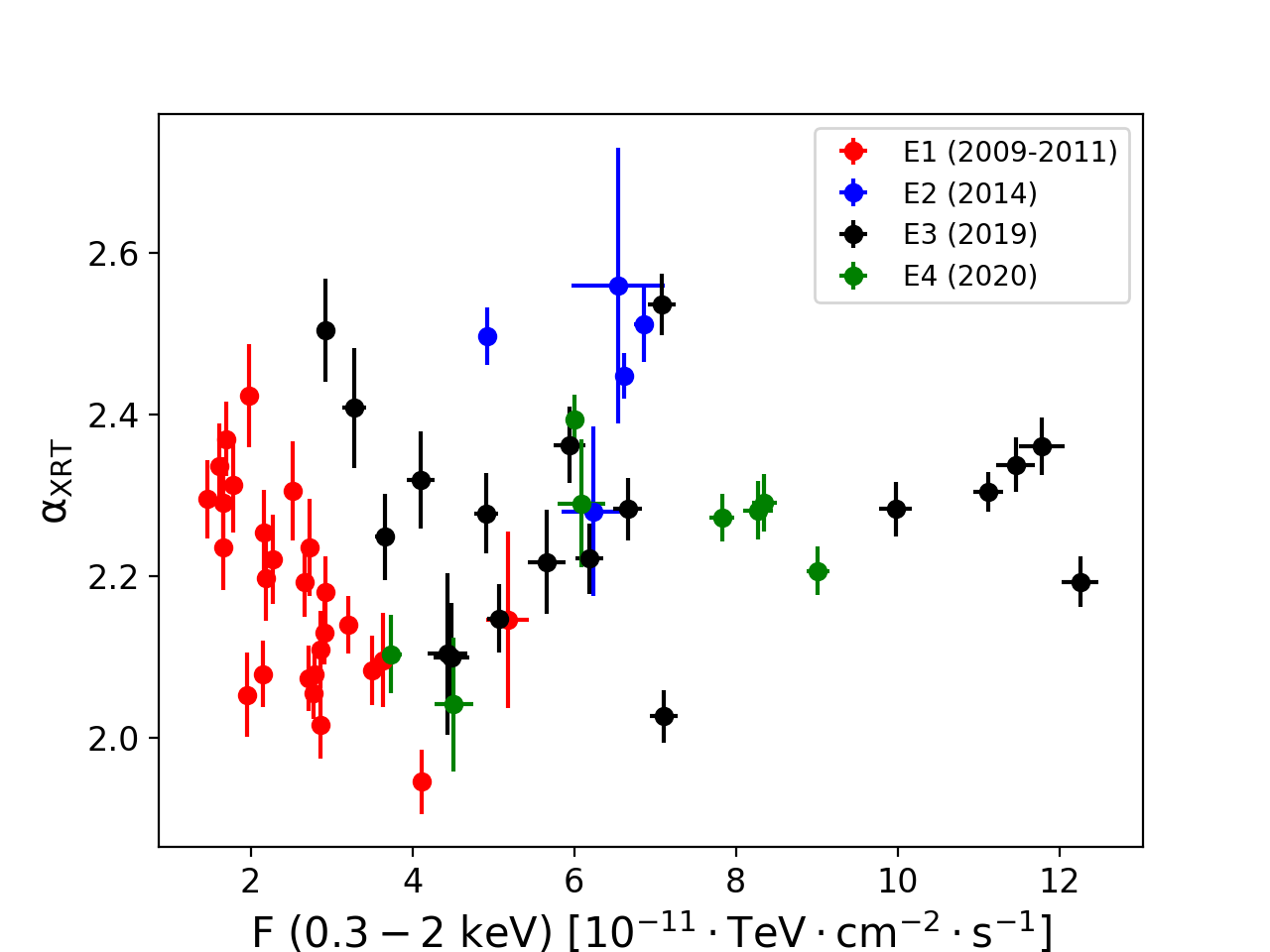}}
\subfigure{\includegraphics[width=0.95\columnwidth]{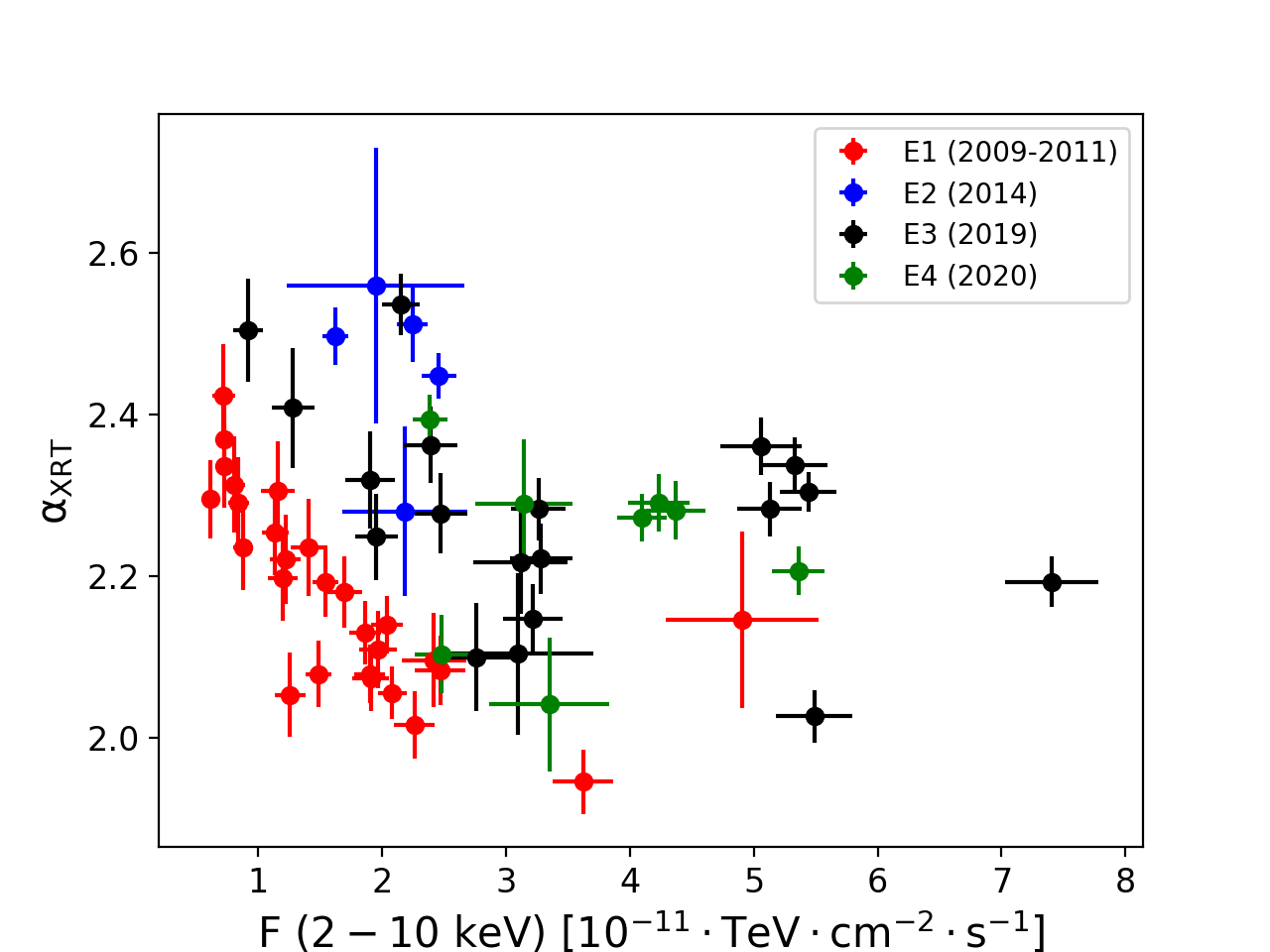}}
\caption{PL index $\alpha_{XRT}$ vs X-ray flux for the different epochs. \textit{Left:} 0.3-2\,keV flux. \textit{Right:} 2-10\,keV flux.}
\label{xray_index_vs_flux}
\end{figure*}

\begin{table*}[]
\caption{Spectral parameters of the X-ray spectrum assuming a simple PL shape.}
\centering
\label{table1}
\begin{tabular}{cccccc}
\hline
Epoch   & \begin{tabular}[c]{@{}c@{}} Time interval \\ {[}MJD{]}  \end{tabular} & \begin{tabular}[c]{@{}c@{}}F$_{0.3-2 \rm~keV}$ \\ {[}10$^{-11}$  $\cdot$ erg $\cdot$ cm$^{-2}$ $\cdot$ s$^{-1}${]} \end{tabular} & \begin{tabular}[c]{@{}c@{}}F$_{2-10 \rm~keV}$ \\ {[}10$^{-11}$ $\cdot$ erg $\cdot$ cm$^{-2}$ $\cdot$ s$^{-1}${]} \end{tabular} & $\alpha_{XRT}$ & $\chi^2$/d.o.f. \\ \hline
\begin{tabular}[c]{@{}c@{}} E1 \\ (Minimum) \end{tabular} & 55516.2  &     1.61$^{+0.06}_{-0.05}$     &    0.73$^{+0.07}_{-0.07}$      &      2.34 $\pm$ 0.05    &          51.0/49 $\simeq$ 1.0  \\[0.35cm] 
\begin{tabular}[c]{@{}c@{}} E1 \\ (Mean) \end{tabular} &  55322.3 -- 55623.8   &    2.26$^{+0.02}_{-0.01}$    &   1.16$^{+0.02}_{-0.02}$      &    2.15 $\pm$ 0.01    &         5.8/5 $\simeq$ 1.2  \\ 
\begin{tabular}[c]{@{}c@{}} E1 \\ (Maximum) \end{tabular}   &   55597.8  &     4.12$^{+0.13}_{-0.16}$     &     3.62$^{+0.27}_{-0.24}$     &    1.95 $\pm$ 0.04     &       66.7/69 $\simeq$ 1.0  \\[0.35cm] 
        E2          &       56987.3         &    4.92$^{+0.11}_{-0.10}$      &    1.62$^{+0.12}_{-0.11}$     &        2.50 $\pm$ 0.04         &    140.9/121 $\simeq$ 1.2     \\[0.35cm] 
        E3     &       58820.2         &     12.3$^{+0.23}_{-0.25}$     &    7.40$^{+0.23}_{-0.25}$    & 2.19 $\pm$ 0.03       &   151.7/153 $\simeq$ 1.0      \\[0.35cm]
\begin{tabular}[c]{@{}c@{}} E4 \\ (Minimum) \end{tabular}     &         59200.0       &      3.73$^{+0.13}_{-0.13}$    &    2.48$^{+0.25}_{-0.21}$    &   2.10 $\pm$ 0.05     &     54.2/47 $\simeq$ 1.2     \\[0.35cm]
\begin{tabular}[c]{@{}c@{}} E4 \\ (Mean) \end{tabular}     &          59196.5 -- 59207.1      &     6.67$^{+0.06}_{-0.06}$     &    3.67$^{+0.09}_{-0.08}$    &    2.27 $\pm$ 0.01    &        6.4/5 $\simeq$ 1.3  \\ 
\begin{tabular}[c]{@{}c@{}} E4 \\ (Maximum) \end{tabular}     &       59202.0         &      9.01$^{+0.16}_{-0.14}$    &    5.36$^{+0.22}_{-0.21}$    &     2.21 $\pm$ 0.03   &    188.5/178 $\simeq$ 1.1      \\ \hline
\end{tabular}
\end{table*}

The different spectra were fitted with a simple PL function ($dN/dE=f_{0}\cdot (E/E_{0})^{-\alpha}$). A LogP fit was also tested ($dN/dE=f_{0}\cdot\,(E/E_{0})^{-\alpha-\beta\,\cdot\,\log(E/E_{0})}$). {Both models also included a photoelectric absorption with a neutral-hydrogen column density fixed to the Galactic value in the direction of 1ES~0647+250, namely $1.20 \times 10^{21}$~cm$^{-2}$ \citep{HI4PI2016}.} However, there is no statistical preference for the LogP over the PL, with $\chi^{2}/\text{d.o.f.}\sim1.0-1.5$ for both models. Thus, for this analysis we assume the simpler spectral shape defined by the PL. The spectra from E1 show a variation of the spectral index of $\sim$0.4, varying from values of $\alpha_{XRT}\sim2.4$ when the source is fainter, up to much harder values of $\alpha_{XRT}\sim2.0$ when the source is in a higher state (see Fig.~\ref{xray_index_vs_flux}). The correlation between the spectral index and the X-ray flux was quantified by estimating the Pearson's linear correlation coefficient for the integral X-ray flux in the 0.3-10 keV band, obtaining a value of $r = -0.65 \pm 0.15$ for this campaign, with a p-value of $3\times10^{-4}$. This behaviour of harder-when-brighter has been seen in the past for other blazars such as 1ES\,2344+514 \citep{acciari2011b,aleksic2013}, Mrk~421 \citep{2021A&A...655A..89M} or TXS~1515-273 during bright X-ray flares \citep{2021MNRAS.507.1528A}. This has found to be a typical behaviour for blazars in the X-ray domain \citep[see e.g.][]{wang2018}.

The spectral analysis of the 2014 flare (E2) reveals a steeper spectrum compared to those from the non-flaring state. During this period, the X-ray spectral index varies around $\alpha_{XRT}\sim2.5$, as can be seen from Fig \ref{xray_index_vs_flux}. We also note that we do not see this harder-when-brighter behaviour during this epoch. However, this could also be due to the sparse time coverage during this period, with only four X-ray observations.

The enhanced state observed in E3 shows the highest X-ray flux ever detected for this source \citep{kapanadze2019}. The spectral index during this flare displays values of $\alpha_{XRT}\sim2.4$, reaching a harder index of $\alpha_{XRT}\sim2.2$ during the maximum of the detection. No clear trend is seen between the spectral index and the X-ray flux for this period. However, a visual inspection of the E3 data set (black points in Fig. \ref{xray_index_vs_flux}) reveals a different behaviour for the faintest measurements (those with a flux $F<8\times 10^{-11}$\,erg\,cm$^{-2}$s$^{-1}$ for the band between 0.3 keV and 2 keV, and $F<4\times 10^{-11}$\,erg\,cm$^{-2}$s$^{-1}$ for the 2-10 keV X-ray band, performed after the VHE flare) than for the brightest measurements (those performed during the historically high X-ray activity and roughly simultaneous to the observations performed by MAGIC). The former subset is characterised by a Pearson's correlation coefficient between the spectral index and the flux for this data set in the 0.3-10 keV band of $r=-0.64 \pm 0.20$ and a p-value of 8$\times10^{-3}$. The latter subset, however, shows no correlation between the flux and the spectral index. This result may indicate a saturation of the X-ray spectral index for the highest X-ray fluxes during the flare observed by \textit{Swift}-XRT.


Finally, during the 2020 observing period (E4) the source displayed significant X-ray variability. This variability was also observed during E1, where the source showed variability both in its X-ray flux (as reported in Table \ref{variability}) and spectral index, with the harder-when-brighter behaviour already reported. The spectral index during E4 ranges from $\alpha_{XRT}\sim2.2$ to $\alpha_{XRT}\sim2.4$. However, contrary to the results from E1, the X-ray observations performed during this period do not reveal any correlated evolution of the spectral index and the flux.

{When the integrated flux in the 0.3-10~keV band was considered, the same results as those presented above were obtained. This suggests that the spectral index does not vary between the 0.3-2~keV and the 2-10~keV X-ray bands}. In summary, two of the epochs (E2 and E4) do not reveal any clear behaviour in their X-ray spectra. On the other hand, a harder-when-brighter behaviour was detected for E1 and E3, in the latter followed by an index saturation for those observations simultaneous to the {brightest X-ray observations, and closest to those from MAGIC}. 
This saturation has also been observed for other blazars like Mrk~421 in the past \citep{2021MNRAS.504.1427A}. {Moreover, the harder-when-brighter trends have been explained in terms of, for instance, a change in the maximum energy of the electrons responsible for the emission \citep[see e.g.][]{abeysekara2017} or a hardening of the electron distribution \citep{xue2006}.}  A detailed discussion on the spectral variability studied here can be found in Sect.~\ref{sec7}. {We also searched for hysteresis loops in the X-ray spectral index evolution during the X-ray flares. However, the low coverage and uncertainties in the spectral index characterisation do not allow us to observe any clear hysteresis-like behaviour.}

\subsection{HE $\gamma$-ray spectral analysis}
The HE $\gamma$-ray spectrum and SED of 1ES\,0647+250 were extracted for all the analysis epochs making use of the \textit{Fermi}-LAT data. A PL shape was assumed. This fit was performed in the energy range of (0.3-300 GeV). The low energy photons (<3-5 GeV, below the IC peak) dominate the fit of the spectrum, leading to the observed `hard' indices. Table~\ref{table:he_sed} shows {the spectral parameters of the HE band} for the different epochs. No significant variability in the slope of the spectrum was detected in the \textit{Fermi}-LAT {observations}. Moreover, contrary to the results for the X-ray spectrum, here we do not see any correlation of the spectral index with the flux. The spectral index light curve is compatible with a constant {value,} $\alpha_{\textit{Fermi}}=1.70 \ \pm \ 0.02$ and {a value of $\chi^2/\text{d.o.f.}=126.8/143 \simeq 0.9$}, which is similar to $\alpha_{\textit{Fermi}}=1.73 \ \pm \ 0.02$ reported in the 4FGL-DR3 catalogue \citep[][]{abdollahi2020,abdollahi2022}.

For the monitoring performed in E1, all the \textit{Fermi}-LAT data from November 2009 to March 2011 were used. For the enhanced states of 2014 (E2) and 2019 (E3), a 12-day integration window centred around the MAGIC detection was used due to the low HE $\gamma$-ray flux displayed by 1ES~0647+250 in {shorter} timescales during these epochs, leading to {a TS<25} and mostly upper limits in the spectral points. Finally, for the data from {2020 (E4)}, \textit{Fermi}-LAT data simultaneous to the MAGIC observations were used to calculate the spectrum. The spectral parameters of each period are shown in Table~\ref{table:he_sed}.

\begin{table*}[]
\centering
\caption{Spectral parameters of the HE $\gamma$-ray spectra from \textit{Fermi}-LAT data.}
\label{table:he_sed}
\begin{adjustbox}{max width=\textwidth}
\begin{tabular}{ccccccc}
\hline
Epoch    & \begin{tabular}[c]{@{}c@{}}Time interval\\ {[}MJD{]}\end{tabular}  & \begin{tabular}[c]{@{}c@{}}$f_{0}$\\ {[}$10^{-7} \cdot$ TeV$^{-1}$ $\cdot$ cm$^{-2}$ $\cdot$ s$^{-1}${]}\end{tabular}  & \begin{tabular}[c]{@{}c@{}}E$_{0}$\\ {[}GeV{]}\end{tabular} & \begin{tabular}[c]{@{}c@{}}Spectral index\\ $\alpha$\end{tabular}  & \begin{tabular}[c]{@{}c@{}}$f \ (0.3-300 \ \text{GeV})$\\ {[}10$^{-9} \cdot$ cm$^{-2}$ $\cdot$ s$^{-1}${]}\end{tabular} & TS \\ \hline 

E1   &   55131.0 -- 55622.0  &  $2.1 \pm 0.2$ & 3  & 1.64 $\pm$ 0.08 &   4.2 $\pm$ 0.6  &  336 \\  

E2   &   56981.0 -- 56993.0  &  $6.1 \pm 2.4$  & 3  & 1.54 $\pm$ 0.25 &   13.9 $\pm$ 6.8  &   41 \\

E3 &   58814.0 -- 58826.0  &  $7.5 \pm 2.4$  & 3  & 1.68 $\pm$ 0.22 &   20.2 $\pm$ 8.7   &  44  \\ 

E4 &   59197.5 -- 59206.5  & $17.8 \pm 4.9$  & 3 & 1.58 $\pm$ 0.17 & 33.0 $\pm$ 0.1 &  84 \\\hline

\end{tabular}
\end{adjustbox}
\tablefoot{The TS value is the likelihood test statistic resulting from the fit to the model.}
\end{table*}

\subsection{VHE $\gamma$-ray spectral analysis}
\label{sec:vhe_sed}
The spectrum and SED were also obtained for the VHE $\gamma$-ray band for the different time periods of the analysis. The spectra from the observations from E1, E2, and E3 were well modelled with PLs. In contrast, a $3\sigma$ preference for a log-parabolic shape was observed in the spectrum of E4. The results of the MAGIC spectral analysis are summarised in Table~\ref{table:vhe_sed}. For the rest of the periods, the statistics are not high enough to evaluate a log-parabolic spectral shape. We note that, since the redshift of the source is still under debate, the spectra and SEDs presented in this analysis correspond to the observed spectra, without the correction for extragalactic background light (EBL) absorption.

\begin{table*}[]
\centering
\caption{Spectral parameters of the VHE $\gamma$-ray spectra from MAGIC data.}
\label{table:vhe_sed}
\begin{adjustbox}{max width=\textwidth}
\begin{tabular}{cccccccc}
\hline
Epoch   &  \begin{tabular}[c]{@{}c@{}}Time interval\\ {[}MJD{]}\end{tabular}  & \begin{tabular}[c]{@{}c@{}}Fit\\ Model$^{*}$ \end{tabular} & \begin{tabular}[c]{@{}c@{}}$f_{0}$\\ {[}10$^{-10} \cdot$ TeV$^{-1}$ $\cdot$ cm$^{-2}$ $\cdot$ s$^{-1}${]}\end{tabular}  & \begin{tabular}[c]{@{}c@{}}E$_{0}$\\ {[}GeV{]}\end{tabular} & \begin{tabular}[c]{@{}c@{}}Spectral index\\ $\alpha$\end{tabular}  & \begin{tabular}[c]{@{}c@{}}Curvature\\ $\beta$\end{tabular} & $\chi^2$/d.o.f. \\ \hline 
E1  &   55131.0 -- 55620.9   & PL &   0.29 $\pm$ 0.07 & 190 & 3.12 $\pm$ 0.37   &   --  &     1.2/3 $\simeq$ 0.4     \\ 
E2   &  56986.2 -- 56987.2    & PL &     4.40 $\pm$ 1.63 &   100 & 3.25 $\pm$ 0.74   &  --   &    2.1/2 $\simeq$ 1.1     \\ 
E3   &   58819.0 -- 58821.2   & PL &     12.0 $\pm$ 2.2 & 100  &  3.73 $\pm$ 0.58    &  --  &    2.2/2 $\simeq$ 1.1      \\ 
E4  &  59198.1 -- 59206.1 & PL & 16.9 $\pm$ 1.0 & 100 & 3.70 $\pm$ 0.10 & -- & 18.1/5 $\simeq$ 3.6 \\
E4  & 59198.1 -- 59206.1 &   LogP & 18.9 $\pm$ 1.6 & 100 & 3.16 $\pm$ 0.21 & 1.91 $\pm$ 0.68 & 5.3/6 $\simeq$ 0.9\\ \hline

\end{tabular}
\end{adjustbox}
\tablefoot{$^{*}$For the 2020 VHE $\gamma$-ray spectrum, both PL and LogP fit models were used, with a ~3$\sigma$ preference for the latter. These functions are specified in Sect. \ref{sec4.1}.}
\end{table*}

\section{Redshift estimation}\label{sec5}
We used the joint \textit{Fermi}-LAT and MAGIC spectra to constrain the redshift of 1ES\,0647+250. To perform this estimation, we followed the procedure proposed by \cite{prandini2010}. This method is based on the assumption that the VHE $\gamma$-ray spectrum of a blazar after correcting for the EBL absorption, cannot be harder than the spectrum in the HE range measured by \textit{Fermi}-LAT. {It makes use of the EBL model developed by \cite{franceschini2008}.} First, an upper limit of the redshift, $z^{*}$, is calculated as the limit value at which the slopes of the HE and VHE $\gamma$-ray spectra are equal {by de-absorbing the MAGIC spectral points until the spectral indices of the HE and VHE spectra are equal}. Then, the empirical formula that relates $z^{*}$ and the reconstructed value of the redshift, $z_{rec}$, with the updated parameters presented in \cite{prandini2011}, is used to estimate the reconstructed value. We performed this estimation with the \textit{Fermi}-LAT and MAGIC SED from E4 (reported in Tables~\ref{table:he_sed} and \ref{table:vhe_sed}) due to their higher flux and thus, higher statistics and lower uncertainties in the flux and spectral index estimation.

This empirical relation applied to our data led to a redshift estimation of $z_{rec} = 0.45 \pm 0.05$. This derived value of the redshift is in agreement with the current lower limit of $z > 0.29$ estimated by \cite{paiano2017} {from spectroscopic observations}, and the most reliable measurement of the distance by \cite{kotilainen2011}, who reported a value of $z=0.41\pm0.06$ {from the detection of the host galaxy}. Moreover, the {maximum redshift} value obtained with this method is $z^{*}=0.75\pm0.11$, compatible with the estimations of the distance of this blazar cited above.

This method has proven to report accurate values of the distance of several blazars in the past, for instance, MAGIC J2001+435 \citep{aleksic2014} or S5 0716+714 \citep{magic2018}. However, it has also reported some inconsistent values, for instance PKS~0447-439, with an estimation of $z=0.20$ by \citet{prandini2012}, later {measured to be} $z=0.343$ by \citet{muriel2015}. We note that this empirical procedure has different caveats and assumptions when estimating the redshift of a source. 
A detailed discussion of these caveats can be found in \cite{prandini2010,prandini2011}, where the redshift of several $\gamma$-ray emitting blazars was properly estimated.

Alternatively, a second upper limit was derived using a maximum likelihood fit of the joint \textit{Fermi}-LAT and the MAGIC SEDs from E4, using a concave LogP as the spectral model, as described in \cite{acciari2019}. Using the EBL model from \cite{dominguez2011}, a scan of redshifts was performed, obtaining a likelihood profile from which the redshift can be constrained. A $15\%$ systematic uncertainty in the overall light throughput was taken into account following the studies in \citet{aleksic2016}. Under these considerations, the 95\% confidence level upper limit for the redshift $z$ is 0.81.
In the following SED modelling we assume the value of $z=0.41$ from \cite{kotilainen2011}, which is in agreement with the value we derived from the empirical relation.


\section{Broadband SED}\label{sec6}
The broadband emission of blazars has been successfully described in the past with models based on leptonic emission processes. However, due to the still arguable origin of the high-energy SED peak, hadronic models have also been found successful {on several occasions}, especially in the scenario of neutrino emission {\citep[see for instance][]{petropoulou2017,kreter2020,petropoulou2020a,petropoulou2020b}.} Each model has its strengths and limitations, with the hadronic models being favoured by uncorrelated optical, X-ray and $\gamma$-ray variability \citep[see e.g.][]{dimitrakoudis2012,bottcher2013}. {Considering the correlated MWL variability observed for this source,} we propose here an interpretation based on leptonic models. While one-component models provide an easier solution due to the smaller number of free parameters, they have not always been adequate to reproduce the emission of $\gamma$-ray blazars {with respect to} two-component models \citep{acciari2020}. 

{We} present the modelling with both one- and two-component leptonic models, comparing their performance and capability to reproduce the observed features. {The modelling was performed assuming the cosmological parameters reported by \citet{planck2020}: a Hubble parameter of $H_{0}=67.4$ km s$^{-1}$ Mpc$^{-1}$, a matter density of $\Omega_m=0.315$ and a dark energy density $\Omega_{\Lambda}=0.685$. The models used here are not time-dependent. Hence, the different epochs are modelled independently. Given the sparse data sets and the large time separations between the different SEDs, no firm conclusions can be drawn on the temporal evolution of the model parameters.}

\subsection{\label{sec:sed_1comp}One-component model}
As an initial approach, a one-component SSC model is used to reproduce the broadband SEDs of 1ES\,0647+250 \citep{tavecchio1998}. This model assumes the existence of a single, homogeneous and spherical emitting region in the jet with size, $R$, Lorentz factor, $\Gamma$, and magnetic field, $B$. The low-energy bump of the SED is due to synchrotron radiation, while the high-energy bump is modelled through SSC. The population of electrons inside the emitting region is assumed to follow a broken {PL} distribution with the Lorentz factor, described by 

\begin{equation}
N(\gamma)=K\gamma^{-n_1}\left(1+\frac{\gamma}{\gamma_{b}}\right)^{n_1-n_2},   \gamma_{min}<\gamma<\gamma_{max}.
\label{broken_PL_electrons}
\end{equation}

\noindent The distribution has a normalisation $K$ between $\gamma_{min}$ and $\gamma_{max}$ and slopes $n_1$ and $n_2$ below and above the break in the electron distribution, $\gamma_{b}$ \citep{maraschi2003}.

The parameters of the one-component models for each epoch are reported in Table~\ref{sedparams}. The resulting models during each VHE $\gamma$-ray detection are shown in Fig.~\ref{SEDmodels}. We scanned several combinations of the parameters described above in order to perform the modelling. The agreement between the model and the data is evaluated through visual inspection of the SEDs shown in Fig. \ref{SEDmodels}. The one-component SSC model is able to describe well the observational data from optical to VHE $\gamma$ rays. However, the radio data {are} strongly self-absorbed and thus, the emission at radio wavelengths is assumed to {originate} from a different region \citep{tavecchio1998}. Therefore, this model is not able to reproduce the radio emission. 

\subsection{\label{sec:sed_2comp}Two-component model}

Alternatively, we also model the SEDs in different epochs with a two-component model based on \citet{tavecchio2011}. This model calculates synchrotron and SSC emission for spherical emission regions while taking into account synchrotron-self absorption.
{The strength of the magnetic field is typically assumed to scale with the distance from the central engine as $d^{-1}$. If the two components are separate, the one responsible for the X-ray and VHE $\gamma$-ray emission is closer to the central engine than the one responsible for radio and optical emission. Therefore, the former needs to have a stronger magnetic field than the latter component, of the order of $\sim$1~G. \citet{tavecchio2016} show that the magnetic field strengths tend to be significantly lower than the values required for equipartition values in one-component models. Moreover, in two-component models it is difficult to reproduce the observed SED with the magnetic field strength values of the order of 1~G. Re-connection layers and radial structures of magnetic fields across the jet are possible ways to invoke reduced local magnetic field strengths \citep[see discussion in][]{2014ApJ...796L...5N}. Therefore, similar} to the approach used in \citet{acciari2020}, we assumed two co-spatial and interacting emission regions to mimic a simple spine-sheath model. The two spherical emission regions are called `core' and `blob', with sizes $R_{\text{core}}>R_{\text{blob}}$. The regions are filled with electrons distributed in Lorentz factor according to a smoothed broken PL (see Eq. \ref{broken_PL_electrons}) and the physical quantities are expressed in the co-moving frame of each individual region. Each of the emission regions has a Lorentz factor, $\Gamma$, size, $R$, and magnetic field strength, $B$. The following constraints were employed to reduce the number of free parameters for this model.

\begin{figure*}
        \includegraphics[width=\textwidth]{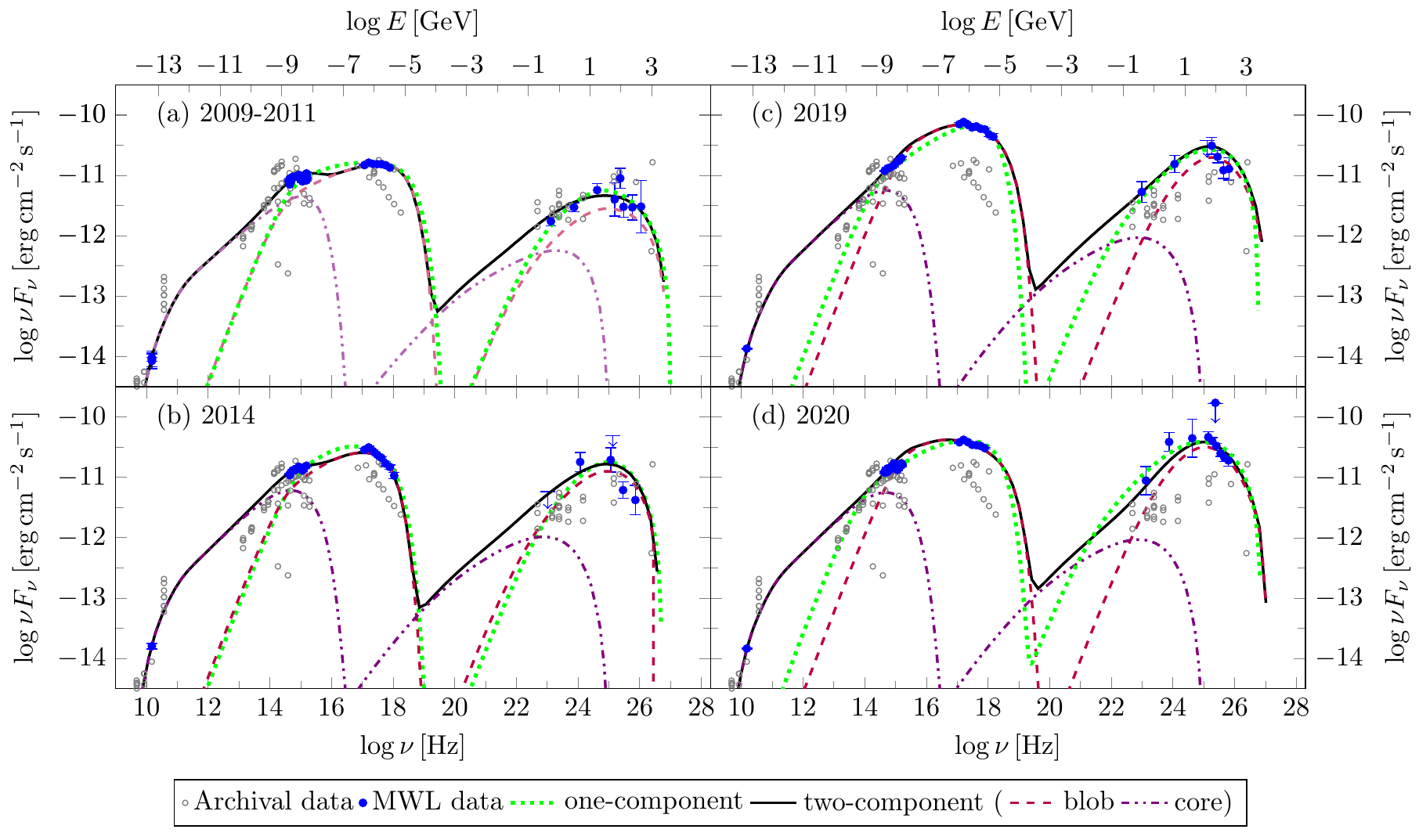}
    \caption{Broadband SEDs of 1ES\,0647+250 for the different observing campaigns described with one-component and two-component SSC emission models. \textit{Top left:} 2009-2011. \textit{Bottom left:} 2014. \textit{Top right:} 2019. \textit{Bottom right:} 2020. Filled blue dots correspond to the MWL simultaneous data of each period. Empty dots represent the archival data extracted from the SSDC database\protect\footnote{\url{https://www.ssdc.asi.it}}. The dotted lines correspond to the fitted one-component SSC model. Black lines represent the total two-component SSC model. The dashed and dotted-dashed lines correspond to the blob and core components of the two-component model, respectively.}
    \label{SEDmodels}
\end{figure*}

\begin{table*}
\caption{\label{sedparams}SED modelling parameters {for one-component SSC and two-component models.}}
\centering  
\renewcommand{\arraystretch}{1.2}

\begin{tabular}{cccccccccccc}
\hline
  (1) & (2) & (3) & (4) & (5) & (6) & (7) & (8) & (9) & (10) & (11) & (12)  \\
  \multirow{2}{*}{Epoch} & Model & $\gamma_{\text{min}}$ & $\gamma_{\text{b}}$ & $\gamma_{\text{max}}$ & \multirow{2}{*}{$n_1$} & \multirow{2}{*}{$n_2$} & B & K & R & \multirow{2}{*}{$\Gamma$} & \multirow{2}{*}{$U'_{B}/U'_{e}$} \\
  
   & (region) & ($\times 10^{3}$) &($\times 10^{4}$) &($\times 10^{5}$) &  &  & (G) & ($\times 10^{3}$\,cm$^{-3}$) & ($\times 10^{15}$\,cm) &  \\
\hline
 \multirow{3}{*}{E1} 
      & one-comp & 5.8 & 2.1 & 6.5 & 2.0 & 3.1 & 0.16 & 2.0 & 34 & 18 & 0.39 \\
      & 2-comp (blob) & 4.5 & 1.9 & 5.5 & 2.0 & 2.9 & 0.16 & 1.0 & 38 & 18 & 0.82 \\
      & 2-comp (core) &  0.2 & 2.2 & 0.4 & 2.0 & 2.4 & 0.16 & 0.04 & 720 & 4 & 12.09\\
\hline

\multirow{3}{*}{E2} 
      & one-comp & 7.0 & 6.3 & 3.4 & 2.02 & 3.6 & 0.16 & 2.5 & 34 & 18 & 0.23 \\
      & 2-comp (blob) & 5.0 & 6.5 & 3.1 & 2.04 & 3.25 & 0.16 & 2.5 & 35 & 17 & 0.27 \\
      & 2-comp (core) & 0.18 & 2.2 & 0.4 & 2.0 & 4.6 & 0.16 & 0.04 & 790 & 4 & 5.96 \\
\hline
  \multirow{3}{*}{E3} 
      & one-comp & 4.0 & 9.4 & 3.9 & 2.07 & 2.9 & 0.18 & 3.9 & 34 & 18 & 0.30 \\
      & 2-comp (blob) & 9.5 & 9.5 & 5.7 & 2.08 & 3.7 & 0.16 & 2.1 & 37 & 23 & 0.49 \\
      & 2-comp (core) & 0.21 & 2.2 & 0.4 & 2.0 & 4.6 & 0.16 & 0.04 & 770 & 4 & 6.14 \\
\hline
\multirow{3}{*}{E4}
      & one-comp & 2.5 & 4.7 & 5.0 & 2.0 & 3.12 & 0.16 & 4.5 & 30 & 17 & 0.09 \\
      & 2-comp (blob) & 9.5 & 5.5 & 6.7 & 2.08 & 3.6 & 0.16 & 7.5 & 29 & 20& 0.16 \\
      & 2-comp (core) & 0.19 & 2.2 & 0.4 & 2.0 & 4.6 & 0.16 & 0.04 & 770 & 4 & 6.02 \\
\hline
\end{tabular}
\tablefoot{Columns: (1) Observation campaign/state. (2) {Model (emission region).} (3), (4) and (5) Minimum, break and maximum electron Lorentz factor. (6) and (7) Slopes of electron distribution below and above $\gamma_{b}$.  (8) Magnetic field strength. (9) Electron density.  (10) Emission-region size. (11) Bulk Lorentz factor. (12) Ratio between the energy density of the magnetic field and the relativistic electrons.}
\end{table*}

First, the measured full-width-half-maximum values of the major axis of {the} very-long-baseline-interferometry core can be used to calculate the upper limit of the size of the core emission region. The measured full-width-half-maximum value of the major axis is 1.88\arcsec\ \citep{piner2014}, which corresponds to $R_{\rm core}\leq 3.3 \times 10^{19}$ cm, assuming a flat universe and $z\simeq0.41$. 

Second, the size of the blob can be constrained from the shortest variability timescales observed as $R\leq \delta c \tau /(1+z)$, where $\delta$ is the Doppler factor and $\tau$ the variability timescale observed.

Third, the bulk Lorentz factor of the core region is limited to 4, which is common for TeV blazars \citep{2018ApJ...853...68P}. The bulk Lorentz factor is then converted to Doppler factor assuming a jet viewing angle $\sim$1/$\Gamma$ and thus $\delta\sim\Gamma$.

Finally, the magnetic field strength of the core can be estimated from the very-long-baseline-interferometry `core shift' measurements \citep{2012A&A...545A.113P} or considering the cooling timescale of the electrons from {the} variability timescale in {the X-ray} band \citep{2018A&A...619A..93B, 2021MNRAS.507.1528A}. Such observations are not available for the source. Therefore, we follow the same assumption employed in \citet{acciari2020} (i.e. $0.1\le B\le 0.4$\,G and similar for core and blob).

The parameters of both the core and blob for the different two-component models of 1ES\,0647+250 are reported in Table \ref{SEDmodels}, and the models are displayed in Fig. \ref{SEDmodels}. We can only set constraints on the size of the blob through the X-ray variability during 2020, with variability {on} timescales as short as 1 day. Considering the minimum and maximum Lorentz factors derived for the blobs, this leads to a blob size $R\leq 3.1 \times 10^{16} - 4.2 \times 10^{16}$~cm. For previous epochs, the variability timescales are longer and thus, the blob size can also adopt higher values. As for the one-component model, the fit is evaluated by visual inspection after scanning the parameters describing the broadband emission. All the two-component models are able to satisfactorily reproduce the broadband emission of our source. The comparison and interpretation of both models with the rest of the results from this study are discussed in Sect.~\ref{sec7}.

\begin{table*}
\caption{\label{sedparams2}Results of the SED modelling {for one-component SSC and two-component models.}}
\centering  
\renewcommand{\arraystretch}{1.2}

\begin{tabular}{cccccccccc}
\hline
  (1) & (2) & (3) & (4) & (5) & (6) & (7) & (8) & (9) & (10)    \\
  \multirow{2}{*}{Epoch} & Model & $L_{\text{B}}$ & $L_{\text{e}}$ & $L_{\text{p}}$ & $L_{\text{jet}}$ & $\log{P_{\text{B}}}$ & $\log{P_{\text{e}}}$ & $\log{P_{\text{p}}}$ & $\log{P_{\text{jet}}}$  \\
  
   & (region) & \multicolumn{4}{c}{($\times 10^{43}$\,erg $\cdot$ s$^{-1}$)} & \multicolumn{4}{c}{(erg s$^{-1}$)}  \\
\hline
 \multirow{2}{*}{E1} 
      & one-comp & 4.79 & 12.3 & 1.06 & 18.1 & 43.55 & 43.96 & 43.96 & 44.34  \\
      & 2-comp (core) & 106.1 & 12.04 & 8.81 & 127.0 & 44.89 & 43.80 & 43.80 & 44.95  \\
\hline

\multirow{2}{*}{E2} 
      & one-comp & 4.79 & 21.2 & 1.30 & 27.3 & 43.55 & 44.20 & 44.20 & 44.55  \\
      & 2-comp (core) &127.8 & 21.49 & 23.55 & 172.8 & 44.97 & 44.19 & 44.19 &  45.09 \\
\hline
  \multirow{2}{*}{E3} 
      & one-comp & 6.07 & 20.1 & 2.09 & 28.3 & 43.66 & 44.18 & 44.18 & 44.54 \\
      & 2-comp (core) & 121.4 & 19.82 & 19.15 & 160.4 & 44.95 & 44.16 & 44.16 & 45.07 \\
\hline
\multirow{2}{*}{E4}
      & one-comp & 3.33 & 36.9 & 5.46 & 45.7 & 43.39 & 44.44 & 44.44 & 44.76  \\
      & 2-comp (core) & 121.4 & 20.1 & 21.19 & 162.7 & 44.95 & 44.17 & 44.17 & 45.07 \\
\hline
\end{tabular}
\tablefoot{Columns: (1) Observation campaign/state. (2) {Model (emission region).} (3), (4), and (5) Kinetic power of the magnetic field, electrons and cold protons, respectively (for the core in the case of the two-component model). (6) Total kinetic power of the jet. (7), (8) and (9) Jet power carried by the jet in form of magnetic field, electrons and cold protons, respectively.  (10) Total power carried by the jet.}
\end{table*}

\section{Discussion}\label{sec7}

\footnotetext{\url{https://www.ssdc.asi.it}}

This manuscript reports the first {detailed} study of the broadband emission {from radio to VHE $\gamma$ rays} of the blazar 1ES~0647+250. The study uses a data set that spans from 2009 to 2020, which allows the multi-band variability and correlations {to be evaluated} over timescales of years. For this, along with the observations performed by the MAGIC telescopes, we make use of radio data from OVRO, several optical telescopes (KVA, LT, LCOGT, and PIRATE), observations performed by \textit{Swift} and its instruments UVOT in the optical and UV regime, and XRT in the X-rays, and HE $\gamma$-ray data from \textit{Fermi}-LAT. In the following subsections we discuss the main results obtained and the implications of the observations reported in previous sections.

\subsection{Variability}
The variability analysis carried out for this source reveals that its broadband emission is clearly variable during the monitored period. The maximum of {$F_{var}$} appears at X-ray and VHE $\gamma$-ray wavelengths, as shown in Fig.~\ref{1ES0647_fractional_variability}. The minimum timescale detected in {these bands} with significant variability {(>5$\sigma$; see Table \ref{variability})} is 1 day in X-rays, only during the flaring state of E4. 
Moreover, when using a 30-day binning for all bands, no significant difference is observed in the structure of $F_{var}$. This is in line with the fact that the variability of this blazar is dominated by the long-term variations of the contribution from {a} common component, as derived by the correlation analysis.

The fact that $F_{var}$ has its maximum in X-rays has been related in the past with a higher dominance of the synchrotron emission \citep[][]{aleksic2015b}. This structure of $F_{var}$ can reveal fundamental differences related to the particle populations and processes producing the broadband emission in blazars. \cite{aleksic2015b} find a similar {$F_{var}$ structure in Mrk~421 to the one shown by 1ES\,0647+250}. One can compare its behaviour with {that} reported for Mrk~501 by \cite{ahnen2017} or \cite{aleksic2015a}, where $F_{var}$ increases with the frequency, reaching its maximum at the VHE $\gamma$-ray band. In the framework of the typical one-component model, the X-ray emission is mainly generated by the high-energy electrons, contrary to the VHE $\gamma$-ray emission, {which is} due to a combination of low-energy and high-energy electrons in Thomson and Klein-Nishina regimes, respectively \citep[][]{abdo2011}. Thus, a higher $F_{var}$ in the X-ray domain, as for the case of 1ES\,0647+250, may be an indication of higher variability of the high-energy electron population, {while the combined low- and high-energy electron distribution may dominate during the $\gamma$-ray flares, which leads to a high $F_{var}$ in VHE $\gamma$ rays.}

\subsection{Correlation and contribution of the two components}
Several studies in the past have found long-term correlations between the optical and radio emission of different sets of blazars with the optical emission leading the radio counterpart {by} a few hundred days \citep[see for instance][]{hufnagel1992,tornikoski1994,hanski2002,ramakrishnan2016,2021MNRAS.504.1427A}. Here, we detect a correlation between the radio and optical emission of 1ES~0647+250 with the optical contribution leading the radio one by 398 days, and between the radio and HE $\gamma$-ray emission, with the $\gamma$ rays {leading} the radio by 393 days. These correlations are detected at a level of $\gtrsim$3$\sigma$, and they are in line with the studies mentioned above. Additionally, a correlation between the optical and $\gamma$-ray emission was found with a time lag compatible with zero. {Due to the long-term nature of the flux variability that leads to the correlations reported here, multi-year data sets are needed to further increase the statistical significance of these results.}

\cite{hufnagel1992} interpreted these correlations as physically related regions, however on different timescales. This has also been stated by other authors \citep[see e.g][]{zhang2017}, and the time lags interpreted as different cooling times between the electrons responsible for the radio and optical emissions \citep{bai2003}. Another plausible explanation is that the radio emission comes from an outer region, but {is} triggered by the same physical mechanisms as the optical and $\gamma$-ray emission. \cite{max-moerbeck2014a} explains this behaviour {as} due to the different opacities for the radio and $\gamma$-ray wavelengths to become observable. Under this scenario, we can estimate the distance between the radio and $\gamma$-ray emitting regions using Eq. 1 from \cite{max-moerbeck2014a}. For this estimation, we use the value of the bulk Lorentz factor $\Gamma=4$ obtained from the core of the two-component SED modelling. {The Doppler factor, $\delta$, is then obtained from the Lorentz factor assuming the approximation of $\delta \sim \Gamma$ for a viewing angle $\theta \sim 1/\Gamma$.}
The redshift value of $z=0.41 \pm 0.06$ from \cite{kotilainen2011} was used. We obtain for a time lag of $-393 \pm 40$ days a distance $d=3.6 \pm 0.4$\,pc. We also estimated the distances for changes on Lorentz factors {by} a factor of 2 ($\Gamma=2$ and $\Gamma=8$) as a conservative comparison of the derived value of $d$ for different values of $\Gamma$, obtaining distances of $d=0.8 \pm 0.1$\,pc and $d=14.9 \pm 1.7$\,pc, respectively. Under this interpretation, the radio and MWL emission would come from physically separated regions. {This estimation can also be performed deriving the Doppler factor as $\delta=[\Gamma\cdot(1-\beta \cos{\theta})]^{-1}$ \citep[see e.g.][]{liodakis2017} instead of the aforementioned small angle approximation. We include this estimation in the Appendix \ref{appendixc} as a comparison.}
{Another plausible interpretation in the scenario of a one-component model is introduced by \cite{tramacere2022}, where the delay between the radio and MWL emissions would be caused by an adiabatic expansion of the emitting region during its propagation along the jet. This expansion leads to a shift of the synchrotron self-absorption frequency to values comparable to or lower than the frequency of the radio emission.}

We also investigated the correlations at shorter timescales, finding no significant correlated emission between {the radio, optical and HE $\gamma$-ray} bands. For this, we performed a detrending of the light curves based on the assumption that the emission is due to a combination of a common and an independent component, following the prescription of \cite{lindfors2016}. We were able to estimate the contribution of this common component to the emission for each pair of light curves. For the radio-optical, radio-$\gamma$-ray and optical-$\gamma$-ray pairs the values found were 0.53, 0.23 and 0.24, respectively. These values are compatible with those reported by \cite{lindfors2016} {for this source}. 

The existence of a common component for the different bands, and the fact that the emission is uncorrelated after subtracting this contribution, may indicate that the long-term variations (timescales of years) are driven by the same mechanism and they come from the common emitting region. In this scenario, the two-component model would be favoured, as the emission would not come from physically separated emitting regions. On the other hand, the short-term variations (timescales of days to months) would be due to the components that do not have common emission. This would be in agreement with the results presented by \cite{ramakrishnan2016}, who state that long-term variations and strong flares are correlated between the radio and optical bands in blazars. Additionally, marginal evidence of correlation between the radio and $\gamma$-ray emission is also reported by these authors. Finally, \cite{ramakrishnan2016} also report strong optical-$\gamma$-ray correlation for a set of blazars, compatible with the results found for 1ES\,0647+250. We note however that this blazar sample is mainly composed of FSRQs, indicating that these correlations may be present in both low- and high-peaking blazars. Additionally, \cite{liodakis2018,liodakis2019} find significant long-term correlations between the radio, optical and $\gamma$-ray bands for more than 100 sources on a sample of 178 blazars, including several HBLs. The optical and $\gamma$-ray emission typically shows time lags compatible with zero, while the radio emission is usually delayed {by} a few hundred days. Moreover, they also find that variability {on} shorter timescales is not necessarily correlated, with the detection of a large fraction of optical and $\gamma$-ray orphan flares.

\subsection{Spectral analysis results}
The spectral analysis carried out on the X-ray data of 1ES\,0647+250 reveals different behaviours for the different observing epochs. During the low state, a clear harder-when-brighter trend between the X-ray flux and the spectral index was observed. X-ray brightening in HBLs has been related in the past to spectral hardening rather than to an overall flux enhancement \citep{giommi2021}, leading to the commonly observed harder-when-brighter evolution for {BL Lac objects} \citep[e.g.][]{pian1998,acciari2011b,aleksic2013,kapanadze2014,balokovic2016,wang2018,2021A&A...655A..89M}. Several conjectures have been proposed to explain such behaviour. A possible explanation could be a hardening or softening of the electron distribution responsible for the synchrotron emission, leading to a hardening (softening) trend with the increasing (decreasing) flux \citep{xue2006}. Additionally, this effect could also be due to an increase in the maximum electron energy or a shift of the synchrotron peak towards higher frequencies \citep{abeysekara2017}. However, this shift of the peak is not evident looking at the different broadband SEDs. 

Moreover, no hardening or softening of the X-ray spectrum was observed during the flares registered. {However, the faintest observations from E3, performed after the MAGIC VHE detection, display again the hardening of the spectrum with the flux increase. On the other hand, the brightest observations and closest to the flare show a rather constant index, which could indicate} the presence of a spectral index saturation during this flare. 
This behaviour is also reported in \cite{2021MNRAS.504.1427A} for the nearby blazar Mrk\,421, {which shows a harder-when-brighter behaviour for a large range of X-ray fluxes, but it shows a sort of saturation for the very high (and very low) X-ray fluxes \citep[][]{2021MNRAS.504.1427A}. In this case, the saturation is present in both the X-ray and VHE bands.}


We also searched for hysteresis processes in the X-ray evolution of the observed flares. These processes have been detected in the past during flares for several blazars, both in X-rays and $\gamma$ rays \citep[see for instance][]{nandikotkur2007,abeysekara2017,wang2018}. These processes provide unique information regarding the acceleration, cooling and energy loss timescales \citep{bottcher2002}, or possible lags between hard and soft X-rays \citep{wang2018}. Nonetheless, no evidence of hysteresis processes was found in the X-ray spectral analysis of 1ES\,0647+250. 

Concerning the $\gamma$-ray band, we do not detect any harder- or softer-when-brighter trend in the long-term variability of 1ES\,0647+250. Despite the fact that the harder-when-brighter trend has been detected in the past also in the HE and VHE $\gamma$-ray domains for some blazars \citep[see e.g.][]{acciari2011a,abeysekara2017,2021A&A...655A..89M}, it has not been systematically observed. Authors like \cite{nandikotkur2007} or \cite{abdo2010} have not found clear correlations between the hardness index and the $\gamma$-ray flux of HBLs. However, we note that the results here may be biased by the wide binning of the \textit{Fermi}-LAT and MAGIC data. 

\subsection{Comparison of one-component and two-component models}
A visual inspection reveals that both models are able to reproduce the MWL emission of this blazar. Both the one-component and two-component models explained the broadband SEDs with a magnetic field of $B=0.16$ G, except {for} the one-component model from E3. This one needs a slightly higher magnetic field ($B=0.18$ G) due to the high dominance of the synchrotron emission with respect to the SSC scattering after the highest X-ray flux detected for this source. In this case, a magnetic field of $B=0.16$~G is unable to well reproduce the broadband emission, leading to large differences between the model and the data regardless {of} the values of the other parameters. These values are in agreement with the typical magnetic fields derived for BL Lacs through a one-component SED modelling by \cite{ghisellini2010}, with $B \sim 0.05-1$ G. Moreover, the Lorentz factor of the one-component model is similar to or lower than that {of} the blob of the two-component, while the core shows a rather small $\Gamma$. \cite{ghisellini2010} infer $\Gamma \sim 10-20$ for a large sample of BL Lacs, compatible with those used in this work to reproduce the broadband emission. The Lorentz factors and the parameters describing the distribution of the electron population $\gamma_{b}$ and $\gamma_{max}$ are also compatible with those derived by these authors.

The one-component model reproduces the broadband emission from the optical regime up to VHE $\gamma$ rays. However, it cannot explain the radio emission with its emitting region. This is in line with the results obtained through the correlation analysis, where the radio was found {to be} delayed { with respect to} the optical and $\gamma$-ray emission. Since the radio photons suffer from strong absorption in the inner regions of the jet, this emission is most likely generated in the outer part of the jet. Thus, the one-component model does not reproduce the radio emission since {this emission} is not co-spatial. 

On the other hand, the two-component can naturally reproduce the emission of 1ES~0647+250 from radio to VHE $\gamma$ rays with co-spatial blob and core regions. In this framework, the core provides seed photons for the IC scattering occurring in the blob. The core dominates the emission from radio to optical wavelengths, while the blob is the main contribution to the emission from X-rays to VHE $\gamma$ rays. This is in line with the results found in \cite{acciari2020}. This leads to higher $\gamma_{min}$ values for the blob, as shown in Table~\ref{sedparams}.

{The use of} a two-component model that reproduces the radio is consistent with the long-term slow trend displayed by 1ES~0647+250. We note that the derived sizes would suggest variability timescales shorter than {those obtained} for the slowly varying component from the data. This means that the slow variability cannot be caused by core-size or acceleration and cooling processes, {which} are generally assumed to be {the} origin of the faster variability; rather, it traces, for example, injection or decay phases of the central engine.

\subsection{Equipartition and jet power}
We calculated the contributions of the magnetic and electron energy densities (see Table~\ref{sedparams}). For the one-component model, \cite{tavecchio2016} show that, for BL Lacs, the magnetic energy density is typically one to two orders of magnitude lower than the electron energy density. Here, we found that, except {for} the model of the 2020 (E4) flare, the parameters suggest the existence of equipartition within a factor of a few. This is in line with the results presented by \cite{nievas2022}, where one-component models are used to reproduce the broadband emission of a sample of HBLs and EHBLs.
\cite{tavecchio2016} also note that by using two-component models it is possible to reproduce the observed SEDs of BL Lacs assuming equipartition between both energy densities. However, other studies based on two-component models \citep[e.g.][for TXS~1515-273]{2021MNRAS.507.1528A} show that, while these models can lead to solutions close to equipartition during low activity states \citep[as reported by][]{tavecchio2016}, {this does not seem to always} be possible, especially during flares.

In the case of two-component models we found that for the core component, the magnetic energy density is dominant over the electron energy density in all cases by an order of magnitude. This ratio for the blob components {is} similar to {that} from {the} one-component model. However, the ratio of the magnetic and electron energy densities is closer to equipartition. This is expected {since} the blob component is much denser {and filled with more-energetic particles} than the core. We calculated the equipartition value for the system of two interacting regions using Eqs. 5, 9, 16, and A1 from \citet{tavecchio2016}. The {ratios} of energy carried out by the magnetic field to {that} by electrons are 38, 0.3, 1.1, and 0.3 for the E1, E2, E3, and E4 data sets. This is in line with the VHE $\gamma$-ray state of the source during {these epochs.} During E1, the source was in a low state when comparing its X-ray and VHE $\gamma$-ray emission with other periods. Therefore, the contribution of energy carried by the electrons is far less than the one carried by the magnetic field. This is reflected in the equipartition values of the whole system as well. It is also evident that during this period, {a} larger amount of emission in the VHE $\gamma$-ray band is coming from the interaction between the components. Such a contribution is smaller for the other periods where the VHE $\gamma$-ray emission is dominated by the emission from the blob.

We also estimated the kinetic energy carried by protons, electrons and the magnetic field of the emitting region. We assumed the simple solution where there is {one} proton per injected electron, and we made use of Eqs. 1 to 3 of \cite{celotti2008}. The results are shown in Table~\ref{sedparams2}, columns (3), (4), and (5). For the one-component model, we find that $L_{e}$ is about one order of magnitude higher than $L_{B}$, as expected from the results reported by \cite{celotti2008} for a much larger population of blazars. This comes from the fact that for BL Lacs, the high-energy emission is of the same order as the synchrotron contribution, and SSC scattering is the only radiative process involved. For the case of {the} core in the two-component models, we found that the kinetic energy carried by protons, electrons and the magnetic field {is} higher than the one obtained in the one-component model.

Finally, we also estimated the power carried by the jet using Eq.~6 from \cite{ghisellini2009}. This estimation was made again under the assumption of one proton per electron and thus, $P_{e} \sim P_{p}$. The estimated values are shown in Table~\ref{sedparams2} along with the total power carried by the jet. The results derived from the one-component model suggest that the power carried by electrons (and protons) tends to be higher than that of the magnetic field. In the context of SSC models, this is a signature characteristic of BL Lacs, where almost all the power of the jet is being used to produce the observed radiation. The values obtained here are also compatible with those estimated for a large sample of BL Lacs, as reported in Fig.~4 from \cite{ghisellini2010}. The values derived from the {two}-component model are also compatible with those reported by these authors. However, in this case, we observed a higher power carried by the magnetic field, in contrast with the lower value for the one-component model.

\section{Conclusions}\label{sec8}

We present, for the first time, a detailed characterisation and interpretation of the broadband emission of the BL Lac object 1ES~0647+250 between the years 2009 and 2020. The main results of this study are summarised as follows:

\begin{itemize}
    \item The MWL emission is clearly variable {on} long timescales, with an increasing flux in radio, optical, and $\gamma$-ray wavelengths. Such behaviour has been seen for other blazars \citep[e.g. 1ES~1215+303;][]{valverde2020}, where the flux increase over year timescales is compatible with that expected from variations in the conditions of the accretion disc. While this interpretation is applicable to the 11-year data set from 1ES~0647+250, more data would be necessary to fully confirm or rule out this hypothesis.
    
    \item A long-term correlation with no delay is measured between the optical and $\gamma$-ray emission at a confidence level of $\sim$3$\sigma$. Moreover, the radio is correlated with the optical and the $\gamma$-ray bands (at a statistical significance of 3$\sigma$ and 4$\sigma$) with time lags of $393 \pm 40$~days and $398 \pm 80$~days, respectively. This time delay is compatible with the radio being emitted from a distinct region of the jet, at a distance of $d=3.6~\pm~0.4$~pc, assuming $\Gamma=4$, $\delta \sim \Gamma$, and $\theta=1/\Gamma$.
    
    \item A harder-when-brighter behaviour in the X-ray spectra is observed during the low state as well as for the flare from 2019 (E3), followed in the latter case by a spectral index saturation. No clear relation is observed in the $\gamma$-ray domain.
    
    \item We estimate the redshift of this object through the comparison of its simultaneous GeV and TeV spectra during a flaring activity \citep[using the method described in][]{prandini2010}, obtaining a value of $z=0.45 \pm 0.05$, which {is} in agreement with some tentative measurements reported in the literature.
    
    \item The broadband SED is characterised and interpreted within {one}- and {two}-component leptonic scenarios for four distinct epochs, namely the low activity in 2009--2011 (E1) and three flaring activities in the years 2014 (E2), 2019 (E3), and 2020 (E4). All the models use a magnetic field of about $B=0.16$~G, and the energy loss of the electron population is dominated by synchrotron emission (needed to match the high X-ray flux and spectra collected during the different epochs). The model parameters required by the two theoretical scenarios to describe the broadband SEDs are similar to those from typical \mbox{BL Lacs} reported in \citet{ghisellini2010} and \citet{tavecchio2016} {for a large sample of objects}.
\end{itemize}

This paper describes a new and comprehensive MWL analysis of the GeV-emitting blazar 1ES~0647+250, one of the few distant gamma-ray blazars detected at VHEs, thoroughly characterising its long-term evolution over the last decade and evaluating its broadband SED for the first time.

\section*{Author contribution}
J. A. Acosta Pulido: Las Cumbres, PIRATE and Liverpool telescope data reduction;
D. Dorner: project management, coordinating MAGIC observations, theoretical interpretation, coordination of MWL data analysis, drafting the manuscript;
V. Fallah Ramazani: modelling, optical data analysis, theoretical interpretation, drafting the manuscript;
T. Hovatta: OVRO data analysis;
S. Kiehlmann: OVRO data analysis;
C. Leto: \textit{Swift} data analysis;
D. Morcuende: VHE $\gamma$-ray data analysis, redshift estimation, drafting the manuscript; 
J. Otero-Santos: VHE $\gamma$-ray data analysis, redshift estimation, modelling, variability and cross-correlation analysis, drafting the manuscript;
D. Paneque: coordinating MAGIC observations, theoretical interpretation, \textit{Fermi}-LAT data analysis, coordination of MWL data analysis, drafting the manuscript;
M. Perri: \textit{Swift} data analysis;
E. Prandini: redshift estimation;
A. C. S. Readhead: OVRO data analysis;
F. Verrecchia: \textit{Swift} data analysis.
The rest of the authors have contributed in one or several of the following ways: design, construction, maintenance and operation of the instrument(s) used to acquire the data; preparation and/or evaluation of the observation proposals; data acquisition, processing, calibration and/or reduction; production of analysis tools and/or related Monte Carlo simulations; discussion and approval of the contents of the draft.

\begin{acknowledgements}
We want to thank the anonymous referee for her/his useful comments and discussion.
We would like to thank the Instituto de Astrof\'{\i}sica de Canarias for the excellent working conditions at the Observatorio del Roque de los Muchachos in La Palma. The financial support of the German BMBF, MPG and HGF; the Italian INFN and INAF; the Swiss National Fund SNF; the grants PID2019-104114RB-C31, PID2019-104114RB-C32, PID2019-104114RB-C33, PID2019-105510GB-C31, PID2019-107847RB-C41, PID2019-107847RB-C42, PID2019-107847RB-C44, PID2019-107988GB-C22 funded by MCIN/AEI/ 10.13039/501100011033; the Indian Department of Atomic Energy; the Japanese ICRR, the University of Tokyo, JSPS, and MEXT; the Bulgarian Ministry of Education and Science, National RI Roadmap Project DO1-400/18.12.2020 and the Academy of Finland grant nr. 320045 is gratefully acknowledged. This work was also been supported by Centros de Excelencia ``Severo Ochoa'' y Unidades ``Mar\'{\i}a de Maeztu'' program of the MCIN/AEI/ 10.13039/501100011033 (SEV-2016-0588, SEV-2017-0709, CEX2019-000920-S, CEX2019-000918-M, MDM-2015-0509-18-2) and by the CERCA institution of the Generalitat de Catalunya; by the Croatian Science Foundation (HrZZ) Project IP-2016-06-9782 and the University of Rijeka Project uniri-prirod-18-48; by the DFG Collaborative Research Centers SFB1491 and SFB876/C3; the Polish National Research Centre grant UMO-2016/22/M/ST9/00382; and by the Brazilian MCTIC, CNPq and FAPERJ.

The \textit{Fermi}-LAT Collaboration acknowledges generous ongoing support from a number of agencies and institutes that have supported both the development and the operation of the LAT as well as scientific data analysis. These include the National Aeronautics and Space Administration and the Department of Energy in the United States, the Commissariat à l’Energie Atomique and the Centre National de la Recherche Scientifique / Institut National de Physique Nucléaire et de Physique des Particules in France, the Agenzia Spaziale Italiana and the Istituto Nazionale di Fisica Nucleare in Italy, the Ministry of Education, Culture, Sports, Science and Technology (MEXT), High Energy Accelerator Research Organization (KEK) and Japan Aerospace Exploration Agency (JAXA) in Japan, and the K. A. Wallenberg Foundation, the Swedish Research Council and the Swedish National Space Board in Sweden. Additional support for science analysis during the operations phase is gratefully acknowledged from the Istituto Nazionale di Astrofisica in Italy and the Centre National d'Etudes Spatiales in France. This work performed in part under DOE Contract DE- AC02-76SF00515.

The authors acknowledge the use of public data from the \textit{Swift} data archive.

This work makes use of observations from the KVA telescope. 

This work makes use of observations from the Liverpool Telescope, operated on the island of La Palma by Liverpool John Moores University in the Spanish Observatorio del Roque de los Muchachos of the Instituto de Astrofisica de Canarias with financial support from the UK Science and Technology Facilities Council. 

This work makes use of observations from the Las Cumbres Observatory global telescope network. 

This article is based on observations made in the Observatorios de Canarias del IAC with the telescope PIRATE operated on the island of  Tenerife by the Open University in the Observatorio del Teide.

This research has made use of data from the OVRO 40-m monitoring program (Richards, J. L. et al. 2011, ApJS, 194, 29), supported by private funding from the California Insitute of Technology and the Max Planck Institute for Radio Astronomy, and by NASA grants NNX08AW31G, NNX11A043G, and NNX14AQ89G and NSF grants AST-0808050 and AST- 1109911.

Part of this work is based on archival data, software, or online services provided by the Space Science Data Centre - ASI.

D.P. acknowledges support from the Deutsche Forschungs gemeinschaft (DFG, German Research Foundation) under Germany’s Excellence Strategy – EXC-2094 – 390783311.

T. H. was supported by the Academy of Finland projects 317383, 320085, 322535, and 345899.

S.K. acknowledges support from the European Research Council (ERC) under the European Unions Horizon 2020 research and innovation programme under grant agreement No.~771282.

W.M. gratefully acknowledges support by the ANID BASAL projects ACE210002 and FB210003, and FONDECYT 11190853.

R.R. acknowledges support from ANID BASAL projects ACE210002 and FB210003, and ANID Fondecyt 1181620.

\end{acknowledgements}

%
%
\bibliographystyle{aa} 
\bibliography{bibliography} 

\begin{appendix} 
\section{MWL light curve description}\label{appendix}
The top panel of Fig.~\ref{1ES0647LCs} shows the 30-day binned MAGIC light curve. The average integral flux above 100\,GeV of the complete data set was estimated to be (2.96 $\pm$ 0.22)~$\times$~10$^{-11}$~cm$^{-2}$s$^{-1}$, which corresponds to (6.2 $\pm$ 0.5)\%~C.U.. During its non-flaring activity, the source showed a VHE $\gamma$-ray flux above 100\,GeV of (0.97~$\pm$~0.24)~$\times$~10$^{-11}$\,cm$^{-2}$s$^{-1}$, equivalent to (2.0~$\pm$~0.5)\%~C.U.. In epoch E2, 1ES\,0647+250 experienced an increase of its $\gamma$-ray emission during its first high activity state, with an integral flux of (1.62~$\pm$~0.78)~$\times$~10$^{-11}$ cm$^{-2}$s$^{-1}$, corresponding to (3.4~$\pm$~1.6)\% of the Crab Nebula flux. Moreover in epoch E3, it displayed another bright state during the enhanced {X-ray activity}, with a flux of (3.82~$\pm$~0.88)~$\times$~10$^{-11}$~cm$^{-2}$s$^{-1}$ (equal to (8.0~$\pm$~1.8)\%~C.U.) above 100\,GeV. The highest VHE $\gamma$-ray emission of this blazar was observed in E4, with a $\gamma$-ray flux above 100\,GeV of (7.10~$\pm$~0.45)~$\times$~10$^{-11}$ cm$^{-2}$s$^{-1}$, which corresponds to (15.0~$\pm$~1.0)\%~C.U..

Regarding the \textit{Fermi}-LAT light curve shown in the second panel of Fig. \ref{1ES0647LCs}, a 30-day binning was also used. The average flux of 1ES\,0647+250 between 300\,MeV and 300\,GeV was estimated to be (0.79\,$\pm$\,0.03)\,$\times$\,10$^{-8}$\,cm$^{-2}$s$^{-1}$. An increasing trend in the flux can be identified in the light curve, reaching a maximum flux of (2.76\,$\pm$\,0.45)\,$\times$\,10$^{-8}$\,cm$^{-2}$s$^{-1}$ at MJD 58115 (2017), a factor of 3.5 times higher than the average flux. During the VHE $\gamma$-ray non-flaring state observed during E1, this blazar also showed its lowest HE $\gamma$-ray emission, with an average flux of (0.43~$\pm$~0.06)~$\times$~10$^{-8}$~cm$^{-2}$s$^{-1}$. Concerning the different enhanced states, the 30-day binning of the light curve does not allow {for the performance of} a comparison of the simultaneous emission between the HE and VHE $\gamma$-ray bands. 

As for the \textit{Swift}-XRT X-ray light curve, the flux in the energy band between 0.3\,keV and 2\,keV ranges from a value of (1.47\,$\pm$\,0.06)\,$\times$\,10$^{-11}$\,erg\,cm$^{-2}$s$^{-1}$ during its minimum up to a historical maximum emission of (12.30\,$\pm$\,0.25)\,$\times$\,10$^{-11}$\,erg\,cm$^{-2}$s$^{-1}$ coincident with the high state detected in E3. The average flux derived from all observations was estimated to be (2.98~$\pm$~0.01)\,$\times$\,10$^{-11}$\,erg\,cm$^{-2}$s$^{-1}$. During the low state observed in epoch E1 the average X-ray flux is (2.22\,$\pm$\,0.02)\,$\times$\,10$^{-11}$\,erg\,cm$^{-2}$s$^{-1}$. Moreover, the quasi-simultaneous XRT observations to the 2014 flare show an emission of (4.92\,$\pm$\,0.11)\,$\times$\,10$^{-11}$\,erg\,cm$^{-2}$s$^{-1}$. The same behaviour is seen in the 2-10\,keV energy band. Finally, the average X-ray emission during the flare from E4 was estimated to be (6.67\,$\pm$\,0.06)\,$\times$\,10$^{-11}$\,erg\,cm$^{-2}$s$^{-1}$.

\textit{Swift}-UVOT observed the source with almost the same temporal coverage as the XRT instrument. This instrument reported the flux of the source over the different periods in three UV and three optical filters. Here, we report the UW1 filter, since it is the one with the highest coverage out of the three UV filters of the instrument. The emission in the UV band ranges from a minimum flux of (0.65\,$\pm$\,0.04)\,mJy during the first campaign up to a maximum of (1.42\,$\pm$\,0.07)\,mJy coincident with the 2019 flare. The average emission during the non-flaring state monitored during E1 was (0.74\,$\pm$\,0.01)\,mJy. The flux increased in E2 up to an average of (1.17\,$\pm$\,0.03)\,mJy, a similar value to the flare from E3, (1.12\,$\pm$\,0.02)\,mJy. The highest average emission was detected during the flare from E4, with an average flux of (1.20\,$\pm$\,0.02)\,mJy.

The R-band optical light curve shows the same long-term increasing trend observed in the HE $\gamma$-ray emission of 1ES\,0647+250. The optical flux ranges from (0.94\,$\pm$\,0.04)\,mJy to (4.41\,$\pm$\,0.07)\,mJy. Between 2009 and 2020, the average optical emission of 1ES\,0647+250 was (1.88\,$\pm$\,0.01)\,mJy. During the VHE $\gamma$-ray low state of epoch E1, the average optical flux was (1.76\,$\pm$\,0.01)\,mJy. The emission of this blazar increased over the years, reaching a flux in the R band of (2.57\,$\pm$\,0.05)\,mJy during the flaring activity of epoch E2. The optical emission reached its maximum in February 2019, and faded down to a flux density of (2.66\,$\pm$\,0.05)\,mJy during the high state detected by MAGIC in E3. Finally, the activity during epoch E4 was similar to the previous period, with an average flux density of (2.72\,$\pm$\,0.01)\,mJy.

The activity in radio wavelengths also displays the same increase seen in HE $\gamma$-ray and optical wavelengths. The flux detected by OVRO goes from a quiescent state of (0.043\,$\pm$\,0.007)\,Jy up to (0.110\,$\pm$\,0.002)\,Jy during its highest emission. The average flux at 15\,GHz during the 11-year time span was estimated to be (0.090\,$\pm$\,0.001)\,Jy. The mean radio emission during the VHE $\gamma$-ray low activity level shows a flux of (0.060\,$\pm$\,0.001)\,Jy. During the flare observed in epoch E2, the most simultaneous OVRO observation to the MAGIC detection reports a flux of (0.106\,$\pm$\,0.012)\,Jy. This corresponds to a factor {of} $\sim$1.8 compared to the average emission during the low state. As for the high state of E3, the measured flux at 15\,GHz was (0.101\,$\pm$\,0.001)\,Jy. Finally, the observation performed by OVRO in epoch E4 reported a flux density of (0.097\,$\pm$\,0.002)\,Jy.

\section{Radio and optical long-term trends}\label{appendixb}
Here we show the long-term trends fitted for the radio and optical R-band light curves of 1ES~0647+250 with the method from \cite{lindfors2016}.

\begin{figure}[h]
\centering
\subfigure{\includegraphics[width=\columnwidth]{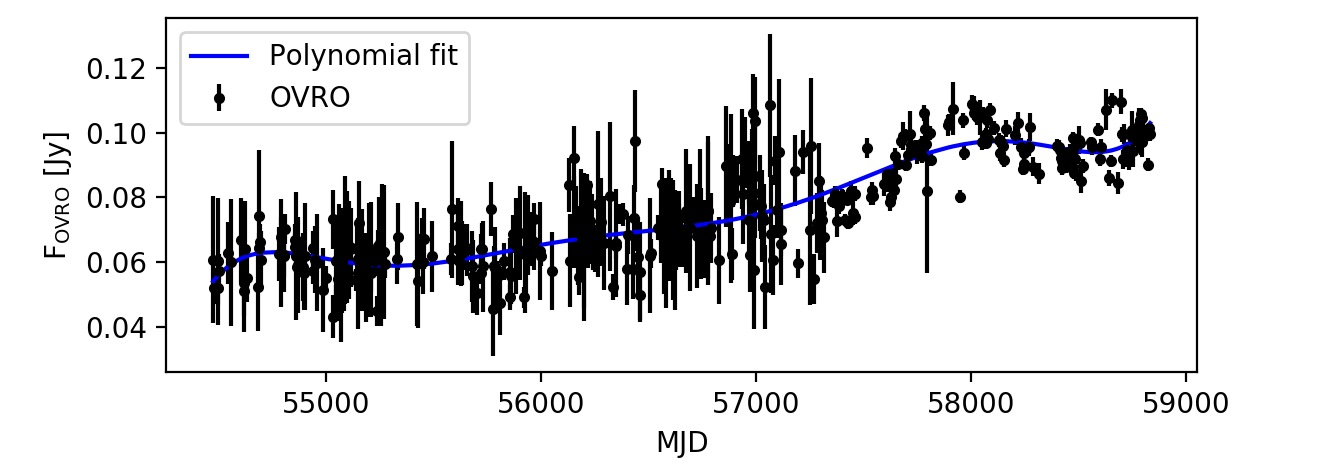}}
\subfigure{\includegraphics[width=\columnwidth]{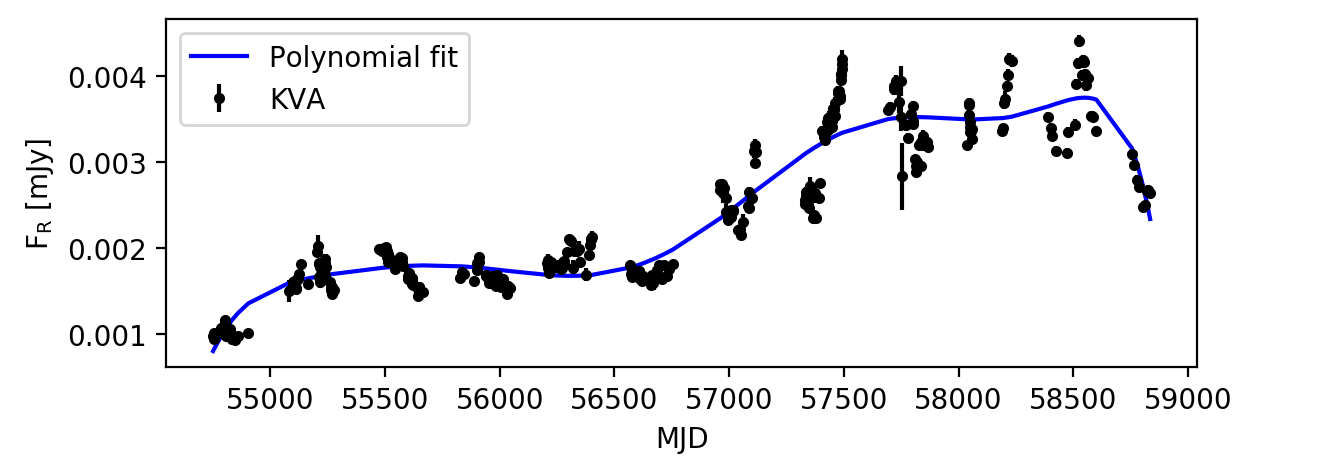}}
\caption{Estimated long-term trends for the 15 GHz radio and R-band optical light curves with the method from \cite{lindfors2016}.}
\label{detrending_fits}
\end{figure}

\section{Estimation of the distance between emitting regions}\label{appendixc}
{
The distance between two physically separated emitting regions can be derived as described in Sect. \ref{sec7}. Instead of using the small angle approximation ($\theta \sim 1/\Gamma$ and therefore $\delta \sim \Gamma$), the Doppler factor can be calculated as}
\begin{equation}
\delta=\frac{1}{\Gamma (1-\beta\cos{\theta})},
\label{doppler_factor}
\end{equation}
where previous information on the viewing angle is needed. In our case there is no estimation on the angle. Therefore, for this calculation we assume a typical value of 1$^{\circ}$. This leads to an estimated distance of $d=1.5 \pm 0.2$\,pc, $d=7.1 \pm 0.8$\,pc, and $d=29.1 \pm 3.4$\,pc for a bulk Lorentz factor of 2, 4, and 8, respectively, following the same approach used in Sect. \ref{sec7}. Therefore, using the Doppler factor resulting of Eq.~\ref{doppler_factor}, with $\theta = 1^{\circ}$ yields a distance larger by approximately a factor of 2 with respect to the $\theta \sim 1/\Gamma$ and $\delta \sim \Gamma$ approximations.

\end{appendix}

\end{document}